\documentclass[a4paper,11pt]{article}
\pdfoutput=1 % if your are submitting a pdflatex (i.e. if you have
\usepackage{jheppub} % for details on the use of the package, please
\usepackage{graphicx}

\usepackage[T1]{fontenc} % if needed

\begin{document}

\title{$B_s\to K^{(*)} \ell\bar \nu$,  Angular Analysis, S-wave Ccontributions  and $|V_{ub}|$}

%% %simple case: 2 authors, same institution
%% \author{A. Uthor}
%% \author{and A. Nother Author}
%% \affiliation{Institution,\\Address, Country}

% more complex case: 4 authors, 3 institutions, 2 footnotes
\author[a,b]{Ulf-G. Mei{\ss}ner}
\author[a]{and  Wei Wang} 
% The "\note" macro will give a warning: "Ignoring empty anchor..."
% you can safely ignore it.

\affiliation[a]{ Helmholtz-Institut f\"ur Strahlen- und Kernphysik and Bethe Center for
Theoretical Physics, Universit\"at Bonn, D-53115 Bonn, Germany }
\affiliation[b]{Institute for Advanced Simulation, Institut f\"ur Kernphysik and J\"ulich
Center for Hadron Physics, JARA-FAME and JARA-HPC, Forschungszentrum J\"ulich,
D-52425 J\"ulich, Germany} 

% e-mail addresses: one for each author, in the same order as the authors
\emailAdd{meissner@hiskp.uni-bonn.de}
\emailAdd{weiwang@hiskp.uni-bonn.de}

\abstract{We analyse the $\overline B_s^0\to K^+l^-\bar \nu$ and $\overline B_s^0\to K^{*+}(\to K\pi)
 \ell^-\bar \nu$ decays that are valuable for extracting  the CKM matrix element  $|V_{ub}|$.  
We calculate the differential and integrated partial widths in units of $|V_{ub}|^2$ based 
on various calculations of hadronic form factors and in particular  the latest Lattice QCD calculation 
of the $B_s\to K^*$ form factors. For the decay $\overline B_s^0  \to K\pi \ell\bar \nu$, 
we formulate the general angular distributions with the inclusion of the various 
partial-wave $K\pi$ contributions.  Using the results for the $K\pi$ scalar form factor 
calculated from unitarized chiral perturbation theory, we  explore the  S-wave effects 
on angular distribution variables and demonstrate that they may not be negligible,  
considering the high precision expected in future measurements. We also briefly discuss 
the impact of the S-wave $\pi\pi$ contributions in the $B^-\to \pi^+\pi^-\ell \bar \nu$ 
decay and provide estimates  for the mode $B^-\to K^+K^-\ell \bar \nu$.  The studies of 
these channels in future can not only be used to determine $|V_{ub}|$, but may also provide 
valuable information  on the $K\pi$ and $\pi\pi$  phase shifts.  }

\maketitle

%%%%%%%%%%%%%%%%%%%%%%%%%%%%%%%%%%%%%%%%%%%%%%%%%%%%%%%%%%%%%%%%%%%%
%\section{Introduction}
%%%%%%%%%%%%%%%%%%%%%%%%%%%%%%%%%%%%%%%%%%%%%%%%%%%%%%%%%%%%%%%%%%%%
\section{Introduction}
  
The precision determination of the Cabibbo-Kobayashi-Maskawa (CKM) matrix element 
$\left| V_{\text{ub}} \right|$ is of particular importance to test the Standard Model 
description of CP  violating effects. Such effects  in weak decays are caused by 
the presence of an irreducible complex phase in the unitary $3 \times 3$ CKM matrix.  
$\left| V_{\text{ub}} \right|$ can be determined from a multitude of  weak $B$-decays
governed by the $b\to u$ transition  which involve either inclusive or exclusive 
final states and exhibit different experimental or theoretical challenges. 
At the current stage,  there is a tension between the   values   extracted
from  inclusive and exclusive  decays. The inclusive determinations
mostly  yield a central value   larger than
$4  \times 10^{-3}$, while    exclusive analyses produce central values 
below this (for a review, See Ref.~\cite{Amhis:2012bh,Beringer:1900zz}). 
Though this tension is only    about 
3$\sigma$, it has already created a significant amount of speculations
about possible new physics effects.

Currently, the process $B\to \pi l\nu_\ell$,  with  $l=e,\mu$,  is considered the most 
reliable exclusive channel to extract $|V_{ub}|$ (for a recent update using  
light-cone sum rules (LCSR) see  Ref.~\cite{Khodjamirian:2011ub}). There is a steady progress in
measuring the branching fraction and $q^2$-distribution on the experimental 
side~\cite{delAmoSanchez:2010af}, while  the theoretical precision is approaching two-loop 
accuracy  in the QCD sum rules~\cite{Bharucha:2012wy}.  On the other hand,  new channels 
that are able to extract  $|V_{ub}|$ and thus can reduce statistical and systematic 
uncertainties  also deserve theoretical and experimental investigations.  
The $\overline B_s^0\to K^+l^-\bar \nu$ and $\overline B_s^0\to K^{*+}\ell^-\bar \nu$ decays 
are of this type. In this work, we shall provide predictions of the differential and 
integrated  decay widths in units of $|V_{ub}|^2$ using the state-of-the-art knowledge 
of the form factors; those include not only the recent Lattice QCD (LQCD) 
calculation~\cite{Horgan:2013hoa} and  the  LCSR~\cite{Ball:2004rg},  
but also various sets of results calculated from the factorisation approach~\cite{Wang:2012ab} 
and QCD-inspired  models~\cite{Faustov:2013ima,Cheng:2003sm,Lu:2007sg,Verma:2011yw}.

Compared to the $B\to \pi \ell\bar \nu$ and $B_s\to K \ell\bar \nu$ decays, 
the $B\to \rho \ell\bar \nu$ and $B_s\to K^* \ell\bar \nu$  reactions receive an additional  
complexity due to the large width of $\rho$ (about 150~MeV) and $K^*$ (about~50 MeV) compared 
to the hadronic scale. 
These processes  are  quasi-four-body decays, and  in
principle   other  $K\pi/\pi\pi$ resonant and nonresonant states may also  contribute
in the same final state, and thus dilute the theoretical  predictions. Another  motivation
of this work is to develop  a general formalism to incorporate various partial-wave
contributions (similar with the $B\to K^*_J(\to K\pi)\ell^+\ell^-$ 
case~\cite{Kruger:1999xa,Lu:2011jm,Doring:2013wka,Li:2010ra,Becirevic:2012dp,Matias:2012qz,Blake:2012mb,Bobeth:2012vn,Descotes-Genon:2013vna,Descotes-Genon:2013wba},   
see also Ref.~\cite{kl4,Kopp:1990yx,Lee:1992ih,Ananthanarayan:2005us,Faller:2013dwa,HKKMM}),  through which the branching ratios, 
forward-backward asymmetries and
polarisations can be   projected out.   In particular, it is worthwhile to  stress  that   
the S-wave, whose effects are not negligible as we will show in the following,   
can {\it not} be expressed    in terms of  a Breit-Wigner formula,  especially for the broad scalar
meson $\kappa~\equiv~K^*_0(800)$. The broad nature is also stressed  from  the Roy-Steiner representations of the $\pi K$ scattering~\cite{Buettiker:2003pp,DescotesGenon:2006uk}.  
To avoid such problem, we will make  use of the Watson 
theorem which allows reliable description in terms of the scalar form factors. These 
scalar form factors have been calculated in dispersion theory or in  unitarization  methods applied 
to  chiral perturbation theory~\cite{Gasser:1990bv,Oller:2000ug,Gardner:2001gc,Meissner:2000bc,Frink:2002ht,Bijnens:2003uy,Lahde:2006wr,Guo:2012yt,Donoghue:1990xh,Jamin:2000wn,Jamin:2001zq,Jamin:2006tj,Bernard:2007tk,Bernard:2009ds}.

Moreover, existing measurements of the branching ratio for the charged current mediated 
$B \to \tau \bar \nu_\tau$ process yield results which are systematically higher than 
the SM expectations~\cite{Lees:2012ju,Adachi:2012mm,Hurth:2012vp}.  The measured decay rate 
for  $B\to D^{(*)}\tau\bar\nu$ is also above the SM  value~\cite{Lees:2012xj,Sibidanov:2013rkk}.  
$B$ decays with $\tau$ leptons in the final state offer possibilities of
significant  new physics (NP) 
contributions not present in processes with light leptons, as
the large $\tau$ mass can overcome the helicity suppression of  certain  decay amplitudes. In this respect, 
the $B_s  \to K^* \tau \bar \nu_\tau$ and $B  \to \rho \tau \bar \nu_\tau$ decays having 
two detectible particles of non-zero spin in the final state ($K^*/\rho, \tau$) offer 
the opportunity of an even more complete investigation of the structure of possible NP 
contributions.  A number of experimental observables sensitive to possible NP effects 
can be introduced.  In the present study, we explore several such observables, like the 
differential distribution over the lepton invariant mass,  the longitudinal $K^*$ branching 
fraction, the  $K^* -\tau$ opening angle distribution, as well as the $\tau$ helicity fractions.

This paper is organised  as follows.  In Sec.~\ref{sec:formfactor}, we give an overview 
of the current knowledge on the heavy-to-light transition form factors. 
The process $\overline B_s\to K^+ \ell \bar\nu$ will be discussed  in Sec.~\ref{sec:BstoK}, 
while the angular distributions for $\overline B_s\to K\pi \ell \bar\nu$ will be derived 
in Sec.~\ref{sec:BstoKstar}.  With these quantities we explore various distribution 
observables for  $\overline B_s^0\to K^0\pi^+ \ell \bar\nu$ and $B^-\to \pi^+\pi^- \ell \bar\nu$ 
including the differential decay widths, the S-wave fraction, and the forward-backward asymmetry. 
We summarise our findings in Sec.~\ref{sec:conclusions}.  Some detailed expressions 
for the angular coefficients are relegated to the Appendix. 
 
 %%%%%%%%%%%%%%%%%%%%%%
\section{Form factors}
\label{sec:formfactor}
 %%%%%%%%%%%%%%%%%%%%%%

After integrating out the virtual $W$-boson,  we arrive at the effective Hamiltonian 
describing the $b\to u$ transition
\begin{eqnarray}
 {\cal H}_{eff} = \frac{G_F}{\sqrt 2} V_{ub} [\bar u \gamma_\mu(1-\gamma_5) b]
[ \bar \ell \gamma^\mu(1-\gamma_5) \nu] +h.c., 
\end{eqnarray}
with $G_F$ the Fermi constant.  The leptonic part is calculable  using  perturbation theory while  
the hadronic effects are encoded in the  transition form factors, 
\begin{eqnarray}
\langle K^+ (p_2) | \bar u  \gamma_{\mu}  b| \overline B_s(p_B)  \rangle
&=& F^{\overline B_s\to K}_{+}(q^2)\left(P_{\mu}-\frac{ m_{B_s}^2-m_{K}^2}{q^2}q_{\mu}
\right)+F^{\overline B_s\to K}_{0}(q^2) \frac{m_{B_s}^2-m_{K}^2}{q^2} q_{\mu}, \nonumber\\
\label{eq:BstoKff}
\end{eqnarray}
with   $q=p_B-p_2$, and $P=p_B+p_2$.  In this $\overline B_s\to K^+$ transition, the strange 
quark serves as a spectator. As we will show in the following,  the parametrization of the 
form factors for higher  resonances with relatively small widths  are also needed in 
the reactions $\overline B_s^0\to K^0\pi^+\ell\bar\nu$.  The decay $\overline
B_s\to K_0^*$, where $K^*_0$ is a scalar resonance, is described by  
\begin{eqnarray}
    \langle K^{*+}_0(p_2) |\bar u  \gamma_\mu\gamma_5  b|
    \overline  {B_s}(p_{B})\rangle
    &=&-i\bigg\{    F_+^{\overline B_s\to K^*_0} (q^2)\left
    [P_\mu - \frac{m_{B_s}^2-m_{K^*_0}^2}{q^2}q_\mu \right ] \nonumber\\
    && 
    +F_0^{\overline B_s\to K^*_0} (q^2)\frac{m_{B_s}^2-m_{K^*_0}^2}{q^2}q_\mu \bigg\},   
  \end{eqnarray}
while the  $\overline B_s\to K_J^*$ form factors with $J \geqslant 1$  are defined
by~\cite{Hatanaka:2009sj,Wang:2010ni,Yang:2010qd}
 \begin{eqnarray}
  \langle K_J^*(p_2,\epsilon)|\bar u\gamma^{\mu}b|\overline B_s(p_B)\rangle
   &=&-\frac{2V(q^2)}{m_{B_s}+m_{K_J^*}}\epsilon^{\mu\nu\rho\sigma} 
   \epsilon^*_{J\nu}  P_{B\rho}P_{2\sigma}, \nonumber\\
  \langle  K_J^*(p_2,\epsilon)|\bar u\gamma^{\mu}\gamma_5 b|\overline
  B_s(p_B)\rangle
   &=&2im_{K_J^*} A_0(q^2)\frac{\epsilon^*_{J } \cdot  q }{ q^2}q^{\mu}
    +i(m_{B_s}+m_{K_J^*})A_1(q^2)\left[ \epsilon^*_{J\mu }
    -\frac{\epsilon^*_{J } \cdot  q }{q^2}q^{\mu} \right] \nonumber\\
    &&-iA_2(q^2)\frac{\epsilon^*_{J} \cdot  q }{  m_{B_s}+m_{K_J^*} }
     \left[ P^{\mu}-\frac{m_{B_s}^2-m_{K_J^*}^2}{q^2}q^{\mu} \right]~.
     \label{eq:BtoTformfactors-definition}
 \end{eqnarray}
We have adopted the convention  $\epsilon^{0123}=+1$, and the polarisation vector 
in the above equations is constructed from the rank $J$
polarisation tensor
\begin{eqnarray}
  &&\epsilon_{J\mu}(h) =\frac{1}{m_{B_s}^{J-1}}
  \epsilon_{\mu\nu_1 \nu_2 
  ...\nu_{J-1}}(h)p_{B_s}^{\nu_1}p_{B_s}^{\nu_2}...p_{B_s}^{\nu_{J-1}},
\end{eqnarray}
with  the helicity $h=0,\pm1$. In the case $J=1$, $\epsilon_{J\mu}=\epsilon_\mu$.

For the $B_s\to K$ transition form factor, no published  result  is available  
from  LQCD simulations, though  some preliminary results can be found in 
Ref.~\cite{Bouchard:2013zda}. The QCD sum rules results  from Ref.~\cite{Duplancic:2008tk} 
do  not provide the analytic information to access  the $q^2$ distribution of the form factors.  
In this work,  our following calculation relies on some QCD-motivated 
models~\cite{Cheng:2003sm,Lu:2007sg,Verma:2011yw,Faustov:2013ima} and the factorisation 
approach~\cite{Wang:2012ab}. For the momentum-transfer distribution, 
the most intuitive and simplest  parametrization of the $B_s\to K$ form factors is the dipole form:
\begin{eqnarray}
 F_i^{B_s\to K}(q^2) &=& \frac{F_i^{B_s\to K}(0)} {(1-a q^2/m_{B_s}^2+ b q^4/m_{B_s}^4)}, \label{eq:dipolePara}
\end{eqnarray}
which has been widely used in the light-front quark model 
(LFQM)~\cite{Cheng:2003sm,Lu:2007sg,Verma:2011yw}. 
In  a recent calculation based on a relativistic quark model (RQM)~\cite{Faustov:2013ima}, 
the $q^2$-dependence of the form factors is  parametrized as 
\begin{eqnarray}
 F_1^{B_s\to K}(q^2) &=& \frac{F_1^{B_s\to K}(0)} {(1-q^2/m_{B}^2)(1-a q^2/m_{B^*}^2+ b q^4/m_{B^*}^4)},\nonumber\\
 F_0^{B_s\to K}(q^2) &=& \frac{F_0^{B_s\to K}(0)} {(1-a q^2/m_{B^*}^2+ b q^4/m_{B^*}^4)}~.
\end{eqnarray} 
The perturbative QCD (PQCD) calculation~\cite{Wang:2012ab} based on the 
$k_T$--factorisation~\cite{Keum:2000ph,Keum:2000wi,Lu:2000em,Lu:2000hj} at next-to-leading order 
in $\alpha_s$~\cite{Li:2012nk}  yields: 
\begin{eqnarray}
 F_1^{B_s\to K}(q^2) &=&  {F_1^{B_s\to K}(0)} \left(\frac{1}{(1-q^2/m_{B_s^*}^2)} + \frac{a q^2/m_{B_s^*}^2}{(1-q^2/m_{B_s^*}^2)(1-b q^2/m_{B_s^*}^2)}\right),\nonumber\\
 F_0^{B_s\to K}(q^2) &=& \frac{F_0^{B_s\to K}(0)} {(1-a q^2/m_{B_s^*}^2+ b q^4/m_{B_s^*}^4)}~.
\end{eqnarray}
Results for these inputs from Refs.~\cite{Faustov:2013ima,Wang:2012ab,Verma:2011yw} are 
collected in Tab.~\ref{tab:BstoKff}, where for the LFQM results, we have also introduced  
the parametric  uncertainties   as done in Ref.~\cite{Chen:2009qk,Wang:2008xt,Wang:2009mi}.

%%%%%%%%%%%%%%%%%%%%%%%%%%%%%%%%%%%%%%%%
\begin{table}%[bth]
\begin{center}
\caption{Theoretical predictions of  the $B_s\to  K$ transitions  form factors, 
in the light-front quark model (LFQM), relativistic quark model (RQM) and the 
perturbative QCD approach (PQCD).  } \label{tab:BstoKff} 
\begin{tabular}{ccccccccc}
     & $F_1(0)$  & $a_{F_1} $ & $b_{F_1} $  & $F_0(0)$  & $a_{F_0} $ & $b_{F_0} $    \\
\hline
 LFQM~\cite{Verma:2011yw}   &$0.23\pm0.01$    &$1.88$  &$1.58$ &$0.23\pm 0.01$  &$1.05$ &$0.35$\\
 PQCD~\cite{Wang:2012ab} & $0.26\pm 0.06$ & $0.57$ & $0.50$ &  $0.26\pm 0.06$ & $0.54$ & $-0.15$   \\
 RQM~\cite{Faustov:2013ima}   &$0.284\pm0.014$    &$-0.37$  &$-1.41$ &$0.284\pm0.014$  &$-0.072$ &$-0.651$  
\end{tabular} 
\end{center}
\end{table}
%%%%%%%%%%%%%%%%%%%%%%%%%%%%%%%%%%%%%%%%%

For $B_s\to K^*$, we use the results from the recent Lattice QCD simulation~\cite{Horgan:2013hoa} 
and  the LCSR~\cite{Ball:2004rg} as the central inputs. 
The LQCD calculation has used the following parametrization to describe the $q^2$-dependence:
\begin{eqnarray}
 F(q^2)=\frac{1}{P(q^2,\Delta m)} [a_0+ a_1z]~, 
\end{eqnarray}
with
\begin{eqnarray}
 z  = \frac{\sqrt{t_+-t}-\sqrt{t_+-t_0}}{\sqrt{t_+-t}+\sqrt{t_+-t_0}}~,
 \nonumber\\
 t_\pm = (m_{B_s}\pm m_{K^*})^2,\;\; t_0 = 12 {\rm GeV}^2~, \nonumber\\
  P(q^2, \Delta m) = 1- \frac{q^2}{(m_{B_s}+\Delta m)^2}~.
\end{eqnarray}
To access the quantity $A_2$, in  the LQCD simulation an auxiliary function
was calculated:   
\begin{eqnarray}
 A_{12}(q^2) = \frac{(m_B+m_V)^2(m_B^2 -m_V^2 -q^2) A_1(q^2) 
- (t_+-t )(t_--t) A_2(q^2) }{16 m_b m_V^2 (m_B+m_V)}~,
\end{eqnarray}
which improves  the numerical  stability. 
In  the LCSR~\cite{Ball:2004rg}, the form factors   are parametrized as 
\begin{eqnarray}
V(q^2)/A_0(q^2) = \frac{r_1}{1-q^2/m_R^2} + \frac{r_2}{1-q^2/m_{\rm fit}^2} \nonumber\\
A_1(q^2) =   \frac{r_2}{1-q^2/m_{\rm fit}^2},  \nonumber\\
A_0(q^2) = \frac{r_1}{1-q^2/m_{\rm fit}^2} + \frac{r_2}{(1-q^2/m_{\rm fit}^2)^2}. 
\end{eqnarray}
The results for these inputs are shown in Tab.~\ref{tab:BstoKstarff}. 
The differences caused by these form factors reflect the 
systematic uncertainties.

%%%%%%%%%%%%%%%%%%%%%%%%%%%%%%%%%%%%%%%%
\begin{table}
\begin{center}
\caption{The $B_s\to  K^*$ transitions  form factors  from the 
LCSR~\cite{Ball:2004rg}  and Lattice QCD~\cite{Horgan:2013hoa}, 
for the $B\to\rho$ transition, we quote the LCSR~\cite{Ball:2004rg} and 
LFQM~\cite{Cheng:2003sm} results.  } \label{tab:BstoKstarff} 
\begin{tabular}{ccccccccccc}
  LQCD~\cite{Horgan:2013hoa}   & $\Delta m$ (MeV)  & $a_0 $ & $a_1 $     \\
\hline 
$V^{B_s \to  {K}^*}$ & $-42$ &       $0.321\pm 0.048$ & $-3.04\pm 0.68$ 
\\
$A_0^{B_s \to  {K}^*}$ & $-87$ &       $0.473\pm 0.042$ & $-2.28\pm 0.74$ 
\\
$A_1^{B_s \to  {K}^*}$ & $350$ &       $0.2337\pm 0.0116$ & $0.082\pm 0.133$ 
\\
$A_{12}^{B_s \to  {K}^*}$ & $350$ &       $0.1919\pm 0.0130$ & $0.376\pm 0.191$ 
\\
\hline
LCSR~\cite{Ball:2004rg} & $F(0)$      & $r_1$ & $m_R^2$ & $r_2$ & $m_{\rm fit}^2$ &      \\ 
\hline
$V^{B_q \to \rho} $&  $0.323 \pm 0.030 $  &    $1.045$ & $5.32^2$ & $-0.721$ & 38.34 &  \\
$A_0^{B_q \to \rho}$ & $0.303 \pm 0.029$ &   $1.527$ & $5.28^2$ & $-1.220 $ & 33.36 &   \\ 
$A_1^{B_q \to \rho}$ & $0.242 \pm 0.023$ &        &           &  $0.240$ & 37.51  \\ 
$A_2^{B_q \to \rho}$ & $0.221 \pm 0.023$ &   $0.009$ &           &$0.212$ & 40.82  \\ 
$V^{B_s \to  {K}^*}$ & $0.311 \pm 0.026$ &       $2.351$ & $5.42^2$ & $-2.039$ & 33.10  \\ 
$A_0^{B_s \to  {K}^*}$ & $0.360 \pm 0.034$ &    $2.813$ & $5.37^2$ & $-2.509$ & 31.58  \\ 
$A_1^{B_s \to  {K}^*}$ & $0.233 \pm 0.022$ &      &   & $ 0.231$ & 32.94  \\ 
$A_2^{B_s \to  {K}^*}$ & $0.181 \pm 0.025$ &    $-0.011$ &   & $0.192$ & 40.14  \\ \hline
  LFQM~\cite{Cheng:2003sm}   & $F_0$  & $a$ & $b$     \\
\hline 
$V^{B \to  \rho}$ & $0.27$ &       $ 1.84$ & $1.28$ 
\\
$A_0^{B \to  \rho}$ & $0.28$ &       $1.73$ & $1.20$ 
\\
$A_1^{B \to  \rho}$ & $0.22$ &       $0.95$ & $0.21$ 
\\
$A_{2}^{B \to  \rho*}$ & $0.20$ &       $1.65$ & $1.05$ 
\\
\hline
\end{tabular} 
\end{center}
\end{table}
%%%%%%%%%%%%%%%%%%%%%%%%%%%%%%%%%%%%%%%%%

In the region where the two pseudo-scalar mesons strongly interact, the resonance approximation 
fails and thus has to be abandoned. One of the such examples is the S-wave   below 
$1$~GeV, for which we can use the form factors as defined  in Ref.~\cite{Doring:2013wka}: 
\begin{eqnarray}
 \langle (K\pi)_0(p_{K\pi})|\bar u \gamma_\mu\gamma_5 b|\overline B_s (p_B)
 \rangle  &=& -i  \frac{1}{m_{K\pi}} \bigg\{ \bigg[P_{\mu}
 -\frac{m_{B_s}^2-m_{K\pi}^2}{q^2} q_\mu \bigg] {\cal F}_{1}^{B_s\to K\pi}(m_{K\pi}^2, q^2) \nonumber\\
 && 
 +\frac{m_{B_s}^2-m_{K\pi}^2}{q^2} q_\mu  {\cal F}_{0}^{B_s\to K\pi}(m_{K\pi}^2, q^2)  \bigg\}~.
 \label{eq:generalized_form_factors}
\end{eqnarray}

Watson's theorem implies that the 
phases measured in $K\pi$ elastic scattering and in a decay channel in which the
$K\pi$ system has no strong interaction with other hadrons  are equal (modulo
$\pi$). 
In the process we consider here, the lepton pair $\ell \bar\nu$ indeed decouple{s from} the 
$K\pi$ final state, and thus we have
\begin{eqnarray}
 \langle (K\pi)_0 |\bar u \Gamma b|\overline B_s\rangle   
 \propto F_{K\pi}(m_{K\pi}^2)~,
\end{eqnarray}
where  the strangeness-changing scalar form factors are defined by
\begin{eqnarray}
 \langle 0| \bar sd |K\pi\rangle = C_X\frac{m_K^2- m_\pi^2} {m_s-m_d} F_{K\pi}( m_{K\pi}^2 )~. 
\label{defff}
\end{eqnarray} 
Here $C_X$ is an isospin factor. In the following, we will consider $K^0 \pi^+$
with $C_X=1$,  and the $K^+\pi^0$ channel is  similar.

It is worthwhile to point out that the generalized form factors also affect charmless 
three-body nonleptonic  $B$--decays under the factorisation assumption, 
see e.g.~\cite{Chen:2002th,Cheng:2013dua,Zhang:2013oqa}.  
An explicit calculation of these quantities~\cite{Meissner:2013hya}. requires the knowledge of generalised 
light-cone distribution amplitudes~\cite{Diehl:2003ny}.  The twist-3 one  has the 
same asymptotic form as the distribution amplitudes  for a scalar resonance~\cite{Cheng:2005nb}.   
Inspired by this similarity, we introduce an intuitive  matching:
\begin{eqnarray}
 {\cal F}_i^{B_s\to K\pi}(m_{K\pi}^2, q^2) = \frac{m_{K}^2-m_\pi^2}{m_s-m_u} \frac{1}{ f_{\kappa}}  F_{K\pi}(m_{K\pi}^2) F_{i}^{B_s\to \kappa} (q^2)~,
\end{eqnarray} 
where the  $ B_s\to \kappa$ form factors have been calculated  in the PQCD 
approach~\cite{Li:2008tk}. 
For the  $B^-\to\pi^+\pi^-\ell\bar\nu$ form factors, we refer the reader to the scalar 
$\pi\pi$ form factors in Refs.~\cite{Oller:1997ti,Lahde:2006wr}  that combine  
unitarization methods  and chiral perturbation theory. Further, we use the
$B\to \sigma$  form factors induced by 
the $b\to u$ transition  from Ref.~\cite{Li:2008tk} which  are collected  in Tab.~\ref{tab:BtoScalar}. Using 
the $B\to f_0(980)$ form factors which are also generated by the $b\to u$
transition,  would enhance the results  on the S-wave contributions given in
the following by  a factor $0.39^2/0.28^2\simeq 1.9$~\cite{Li:2008tk}.

%%%%%%%%%%%%%%%%%%%%%%%%%%%%%%%%%%%%%%%%
\begin{table}%[bth]
\begin{center}
\caption{Theoretical results for  the $B_s\to  \kappa$  and $B\to\sigma$  form factors 
in the perturbative QCD approach~\cite{Li:2008tk} with the parametrization given
in Eq.~\eqref{eq:dipolePara}.  } \label{tab:BtoScalar} 
\begin{tabular}{ccccccccc}
     & $F_1(0)$  & $a_{F_1} $ & $b_{F_1} $  & $F_0(0)$  & $a_{F_0} $ & $b_{F_0} $    \\
\hline
 $B_s\to\kappa$   &$0.29\pm0.07$    &$1.62$  &$0.56$ &$0.29\pm0.07$  &$1.68$ &$0.62$  \\
 $B\to\sigma$  &$0.28\pm0.07$    &$1.61$  &$0.56$ &$0.28\pm0.07$  &$0.65$ &$-0.11$  
\end{tabular}
\end{center} 
\end{table}
%%%%%%%%%%%%%%%%%%%%%%%%%%%%%%%%%%%%%%%%%

%%%%%%%%%%%%%%%%%%%%%%%%
%\section{Angular distributions analysis}
%\label{sec:angularAA}
%%%%%%%%%%%%%%%%%%%%%%%%

%%%%%%%%%%%%%%%%%%%%%%%%
\section{{\boldmath$B_s\to K\ell \bar \nu_{\ell}$}}
\label{sec:BstoK}
%%%%%%%%%%%%%%%%%%%%%%%%

%The kinematics  is given in Fig.~\ref{fig:kinematicsK}. 

%%%%%%%%%%%%%%%%%%%%%%%%%%%%%%%%%%%%%%%%%%%%%%%%%%%%%%%%%%%%%%%%%%%%%%%%%%%%%%%%
%\begin{figure}\begin{center}
%\includegraphics[scale=0.5]{three-body.eps} 
%\caption{Kinematics } \label{fig:kinematicsK}
%\end{center}
%\end{figure}
%%%%%%%%%%%%%%%%%%%%%%%%%%%%%%%%%%%%%%%%%%%%%%%%%%%%%%%%%%%%%%%%%%%%%%%%%%%%%%%% 
 
\begin{figure}\begin{center}
\includegraphics[scale=0.5]{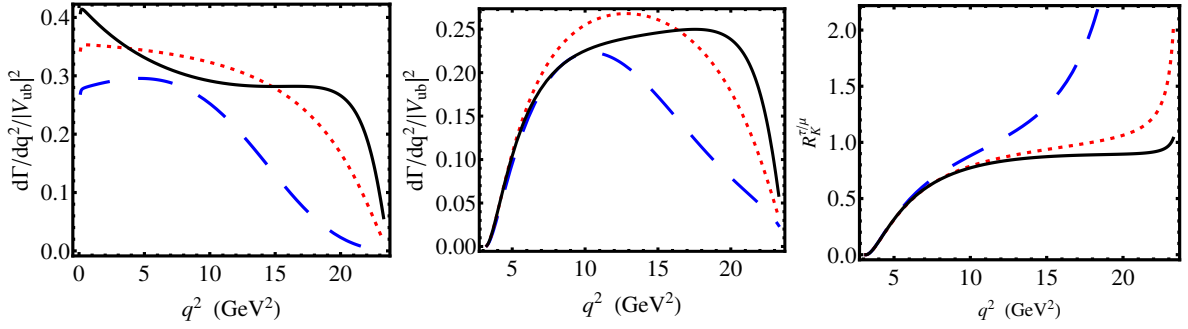} 
\caption{ Differential decay widths (in units of $1/({\rm ps} \times {\rm GeV}^2)$) for 
$\overline B_s^0\to K^+ \ell \bar\nu_{\ell}$ with $\ell= \mu,e$ in the first  panel  and 
$\ell=\tau$ in the second  panel. The $q^2$-dependent ratio $R_{K}^{\tau/\mu}$ defined in 
Eq.~\eqref{eq:RKtauOvermu} is given in the last panel. The solid, dashed and 
dotted curves correspond to the form factors calculated using the RQM, LFQM and PQCD 
approachess, respectively.  } \label{fig:diffdwBsK}
\end{center}
\end{figure}

With the form factors defined in Eq.~\eqref{eq:BstoKff}, 
we evaluate the $B_s\to K$ matrix elements   as
\begin{eqnarray}
H_0 = \frac{G_{F}}{\sqrt 2} V_{ub} \frac{\sqrt{\lambda}} {\sqrt{q^2} } F_{1}^{B_s\to K}(q^2) ,\;\; 
H_t = \frac{G_{F}}{\sqrt 2} V_{ub} \frac{m_{B_s}^2-m_{K}^2} {\sqrt{q^2} }  F_{0}^{B_s\to K}(q^2), 
\end{eqnarray}
and obtain the
differential decay rate for  $\overline B_s^0\to K^+  \ell \bar\nu_{\ell}$ as a
function of $q^2$ and $\theta_l$ 
\begin{eqnarray}
\frac{d^2\Gamma(\overline B_s\to K^+ \ell^- \bar \nu_{\ell})}{dq^2 d\cos\theta_l}&=&
\frac{1}{ 512\pi^3 m_{B_s}^3} \sqrt{\lambda}  4q^2 \beta_l^2   [\sin^2\theta_l |H_{0}|^2 + \hat m_l^2 |H_{t}+H_0\cos\theta_l|^2],
\label{eq:dgammadq2dthetaBstoK}
\end{eqnarray}
where $\beta_l =1- \hat m_l^2$,  $\hat m_{i}=m_{i}/\sqrt{q^2} $ and 
$\lambda = \lambda(m_{B_s}^2, q^2, m_{K}^2) $ is the K\"allen function
\begin{eqnarray}
 \lambda (a,b,c) = a^2+ b^2 +c^2- 2(ab+bc+ca)~. 
\end{eqnarray}
Here, $\theta_l$ is defined as the polar angle of the lepton momentum
relative to the moving direction of the $B_s$-meson in the $q^2$ rest
frame.  Integrating over the polar angle, the differential partial width in $q^2$
is given by
\begin{eqnarray}
\frac{d\Gamma(\overline B_s\to K^+ \ell^- \bar \nu_{\ell})}{dq^2 }&=&
\frac{1}{ 128\pi^3 m_{B_s}^3} \sqrt{\lambda}   q^2 \beta_l^2   \left[\frac{4}{3}  |H_{0}|^2 + 2 \hat m_l^2 |H_{t}|^2+\frac{2}{3}\hat m_l^2|H_0|^2\right]~.
\end{eqnarray}
One can also explore the $q^2$-dependent ratio 
\begin{eqnarray}
{\cal R}_{K}^{\tau/\mu}(q^2) = \frac{ {d\Gamma(\overline B_s\to K^+ \tau^- \bar \nu_{\tau})}/{dq^2 }}{ {d\Gamma(\overline B_s\to K^+ \mu^- \bar \nu_{\mu})}/{dq^2 }}, \label{eq:RKtauOvermu}
\end{eqnarray}
and its integrated form:
\begin{eqnarray}
 R_{K}^{\tau/\mu} = \frac{ {\Gamma(\overline B_s\to K^+ \tau^- \bar \nu_{\tau})}}{ {\Gamma(\overline B_s\to K^+ \mu^- \bar \nu_{\mu})}}, \label{eq:RKtauOvermuNorm}
\end{eqnarray}
where  the $\mu$ lepton can also be replaced by the electron.  
Using three sets of form factors, we show the results for the 
$\overline B_s\to K^+ \ell^- \bar \nu_{\ell}$ differential decay widths 
$d\Gamma/dq^2/|V_{ub}|^2$  (in units of  $1/({\rm ps} \times {\rm GeV}^2)$) 
in Fig.~\ref{fig:diffdwBsK},  with $\ell= \mu,e$ in the first panel  and $\ell=\tau$ in the second
panel. The  ratio $R_{K}^{\tau/\mu}(q^2)$   is also  given in the last panel.  
The solid, dashed and dotted curves correspond to the form factors calculated 
from the RQM, LFQM and PQCD approaches.   Errors  caused by the input parameters 
in the $R_{K}^{\tau/\mu}(q^2)$ 
mostly  cancel  in the three individual sets of calculations, but 
distributions are quite different especially  in the large $q^2$ region.

%%%%%%%%%%%%%%%%%%%%%%%%%%%%%%%%%%%%%%%%%%%%%%%%%%%%%%%%%%%%%%%%%%%%%%%%%%%%%%%%
%%%%%%%%%%%%%%%%%%%%%%%%%%%%%%%%%%%%%%%%%%%%%%%%%%%%%%%%%%%%%%%%%%%%%%%%%%%%%%%% 

%%%%%%%%%%%%%%%%%%%%%%%%%%%%%%%%%%%%%%%%%%%%%%%%%%%%%%%%%%%%%%%%%%%%%%%%%%%%%%%%
\begin{figure}\begin{center}
\includegraphics[scale=0.6]{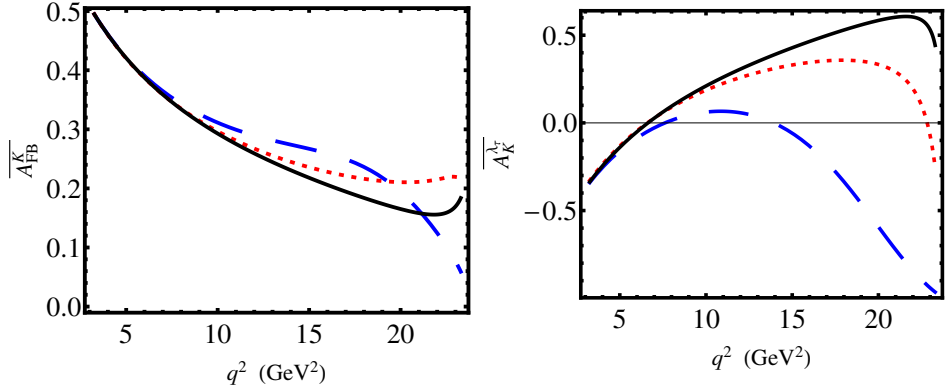} 
\caption{Same as Fig.~\ref{fig:diffdwBsK} but for the forward-backward 
asymmetry (the left panel) defined in Eq.~\eqref{eq:AFBBstoK},  and the  
polarisation fraction  for the $\tau$ lepton (the right panel).  } \label{fig:asymmetry}
\end{center}
\end{figure}
%%%%%%%%%%%%%%%%%%%%%%%%%%%%%%%%%%%%%%%%%%%%%%%%%%%%%%%%%%%%%%%%%%%%%%%%%%%%%%%% 

%%%%%%%%%%%%%%%%%%%%%%%%%%%%%%%%%%%%%%%%%%%%%%%%%%%%%%%%%%%%%%%%%%%%%%%%%%%%%%%%
% \input{tableBsK.tex}
\begin{table}[t]
\begin{center}
\caption{Integrated decay widths for
 $B_s\to K^+ \ell\bar  \nu$:
 $\Delta \zeta_K^{\ell} (q_l^2, q_u^2)$. 
Results are given in units of $ps^{-1}$.}
\label{tab:integratedWidthBstoK}
\begin{tabular}{cccccc}
\hline\hline
& LFQM & PQCD & RQM \\\hline
$\Delta \zeta_K^\mu(0, 4)$
 & $1.12\pm 0.099 $&$ 1.36\pm 0.703 $&$ 1.47\pm 0.149 $ 
\\
 $\Delta \zeta_K^\mu(4, 8)$ 
 & $1.17\pm 0.099 $&$ 1.36\pm 0.702 $&$ 1.28\pm 0.135 $ 
\\
$\Delta \zeta_K^\mu(8, 12)$
 & $1.\pm 0.089 $&$ 1.29\pm 0.664 $&$ 1.17\pm 0.116 $ 
\\
$\Delta \zeta_K^\mu(0, 23.77)$
 & $4.18\pm 0.372 $&$ 6.38\pm 3.29 $&$ 6.83\pm 0.688 $ 
\\
\hline
 $\Delta \zeta_K^\tau(m_\tau^2, 8)$
 & $0.551\pm 0.049 $&$ 0.614\pm 0.317 $&$ 0.577\pm 0.058 $ 
\\
$\Delta \zeta_K^\tau(8, 12)$
 & $0.872\pm 0.077 $&$ 1.01\pm 0.516 $&$ 0.894\pm 0.09 $ 
\\
 $\Delta \zeta_K^\tau(m_\tau^2, 23.77)$
 & $2.82\pm 0.25 $&$ 3.95\pm 2.03 $&$ 4.04\pm 0.406 $ 
\\
\hline
$ R_{K}^{\tau/\mu} $
 & 0.675 & 0.619 & 0.592
\\
 \hline\hline
\end{tabular} 
\end{center}\end{table}

%%%%%%%%%%%%%%%%%%%%%%%%%%%%%%%%%%%%%%%%%%%%%%%%%%%%%%%%%%%%%%%%%%%%%%%%%%%%%%%%

Since the differential decay rate in Eq.~\eqref{eq:dgammadq2dthetaBstoK} involves
the polar angle of the lepton,   we can define an angular asymmetry: 
\begin{eqnarray}
{\cal A}_{FB}^K(q^2)=\left[\int^{1}_{0}-\int^{0}_{-1}\right]d\cos\theta_l  \frac{ d^2\Gamma(\overline B_s\to K^+ \ell^- \bar \nu_{\ell})}{dq^2d\cos\theta_l }. \label{eq:AFBBstoK}
 \label{eq:asy}
\end{eqnarray}
More explicitly, the normalised  asymmetry for $\overline B_s\to K \ell
\bar \nu_{\ell}$ decay is given by
\begin{eqnarray}
\overline {\cal A}_{FB}^K(q^2)= \frac{{\cal A}_{FB}^K(q^2)}{d\Gamma/dq^2}= \frac{2\hat m_l^2 {\rm Re}[H_0 H_{t}^*] }{ (4/3+2/3 \hat m_l^2) |H_0|^2 + 2\hat m_l^2 |H_t|^2}  . \label{eq:asy_P}
\end{eqnarray}
Clearly, the angular asymmetry is only associated with the ratio of
form factors, which supposedly is  less sensitive to the model-dependent
hadronic parameters. Therefore, this quantity could be a
good candidate to explore the new physics effects.

The lepton is produced from the $V-A$ current in the SM, and thus the lepton and muon is 
mainly left-handed polarised. For the   $\tau$ lepton, we can also explore  
the polarised distribution, 
\begin{eqnarray}
\frac{d^2\Gamma(\overline B_s\to K^+ \tau^- \bar \nu_{\ell}(\lambda_\tau=-1/2))}{dq^2 d\cos\theta_l}&=&
\frac{1}{ 128\pi^3 m_{B_s}^3} \sqrt{\lambda}   q^2 \beta_l^2    \sin^2\theta |H_{0}|^2  ,\nonumber\\
\frac{d^2\Gamma(\overline B_s\to K^+ \tau^- \bar \nu_{\ell}(\lambda_\tau=1/2))}{dq^2 d\cos\theta_l}&=&
\frac{1}{ 128\pi^3 m_{B_s}^3} \sqrt{\lambda}   q^2 \beta_l^2   \hat m_l^2 |H_{t}+H_0\cos\theta|^2, 
\end{eqnarray}
and   the polarisation fraction: 
\begin{eqnarray}
\overline  A_{K}^{\lambda_\tau}(q^2) &=&  \frac{ {d\Gamma(\overline B_s\to K^+ \tau^- \bar \nu_{\tau})(\lambda_\tau=-1/2)}/{dq^2 }- {d\Gamma(\overline B_s\to K^+ \tau^- \bar \nu_{\tau})(\lambda_\tau=1/2)}/{dq^2 }}{ {d\Gamma(\overline B_s\to K^+ \tau^- \bar \nu_{\tau})}/{dq^2 }} \nonumber\\
&=& \frac{ (4/3 -2/3 \hat m_l^2) |H_0|^2 - 2\hat m_l^2 |H_t|^2}{ (4/3+2/3 \hat m_l^2) |H_0|^2 + 2\hat m_l^2 |H_t|^2}~. 
\end{eqnarray}
Results for the   asymmetries and polarisations  are given in Fig.~\ref{fig:asymmetry}, 
where different theoretical calculations of form factors lead to different 
behaviours  especially in the large $q^2$ (low recoil) region. This can be improved once  
the Lattice QCD can constrain the form factors.  In this procedure,
we also notice that   Lattice QCD simulation often requests an extrapolation from the unrealistic quark mass region  to physical region. The  hard pion chiral perturbation theory approach advocated in Refs.~\cite{Bijnens:2010ws,Bijnens:2010jg}  has the advantage to resum the chiral logarithms and thus are valuable for the extrapolation. Such effects should be taken into account in the future Lattice QCD simulation.

The integrated decay widths in terms of $|V_{ub}|^2$ are given in 
Tab.~\ref{tab:integratedWidthBstoK}, with the definition  
\begin{eqnarray}
 \Delta \zeta_K^{\ell} (q_l^2, q_u^2) =\frac{1}{|V_{ub}|^2}  \int_{q_l^2}^{q_u^2} dq^2 \frac{d\Gamma(\overline B_s\to K^+\ell \bar\nu)}{dq^2}~. 
\end{eqnarray}
These values will be useful to extract the $|V_{ub}|$ when compared to the   
experimental data available in future.  
The PQCD approach gives the largest errors due to the uncertainties in the form factors, 
while the other approaches have approximately $10\%$ parametric  errors.

%%%%%%%%%%%%%%%%%%%%%%%%%%%%%%%%%%%%%%%%%%%%%%%%%%%%%%%%%%%%%%%%%%%%%%%%%%%%%%%%
\section{Full angular distribution of {\boldmath$B_s\to K\pi  \ell \bar \nu$}}
\label{sec:BstoKstar}
%%%%%%%%%%%%%%%%%%%%%%%%%%%%%%%%%%%%%%%%%%%%%%%%%%%%%%%%%%%%%%%%%%%%%%%%%%%%%%%%

%%%%%%%%%%%%%%%%%%%%%%%%%%%%%%%%%%%%%%%%%%%%%%%%%%%%%%%%%%%%%%%%%%%%%%%%%%%%%%%%
\begin{figure}\begin{center}
\includegraphics[scale=0.6]{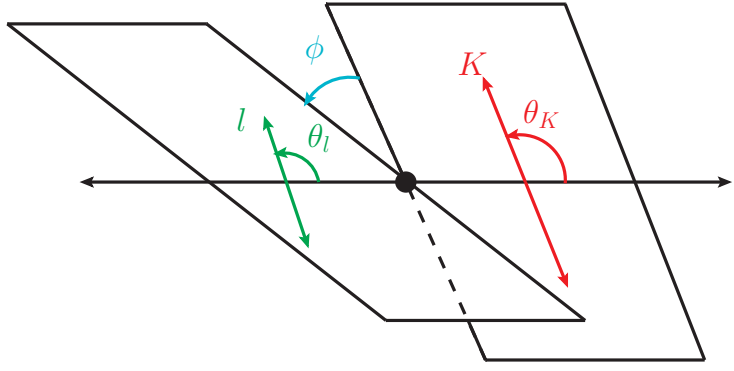} 
\caption{Kinematics for $\overline B_s\to K\pi  \ell \bar \nu$. {The} $K\pi $ system 
moves along the $z$ axis in the $\overline B_s$ rest frame. 
$\theta_K(\theta_l)$ is defined in the $K\pi$ (lepton pair) rest-frame as the angle
between the $z$-axis and the direction of motion of the $K $ ($\ell^-$), respectively.  
The azimuth angle $\phi$ is the relative  angle between the $K\pi$ decay and lepton pair
planes. } \label{fig:kinematicsKpi}
\end{center}
\end{figure}
%%%%%%%%%%%%%%%%%%%%%%%%%%%%%%%%%%%%%%%%%%%%%%%%%%%%%%%%%%%%%%%%%%%%%%%%%%%%%%%% 

We consider the kinematics  for the $B_s\to K\pi  \ell \bar \nu$ as shown in 
Fig.~\ref{fig:kinematicsKpi}.  {The} $K\pi $ system  moves along the $z$-axis 
in the $\overline B_s$ rest-frame. 
$\theta_K(\theta_l)$ is defined in the $K\pi$ (lepton pair) rest frame as the angle
between $z$-axis and the direction of motion of the $K $ ($\ell^-$), respectively.  
The azimuth angle $\phi$ is the relative angle between the $K\pi$ decay and lepton pair
planes.

The decay amplitudes for $B_s\to   K\pi \ell \bar \nu_{\ell}$ can be 
divided into  several individual pieces and each of them can be expressed 
in terms of the  Lorentz invariant helicity amplitudes.  
The amplitude for the hadronic part can be obtained from  the matrix element: 
\begin{eqnarray}
A_{\lambda}  =   \sqrt{ N_{K_J^*}}   \frac{iG_{F}}{\sqrt 2} V_{ub} 
\epsilon_{\mu}^*(h) \langle K\pi |\bar u \gamma^\mu(1-\gamma_5) b |\overline B_s\rangle,
\end{eqnarray} 
where $\epsilon_\mu(h)$ is an auxiliary polarisation vector for the 
lepton pair system, $h= 0, \pm, t$, and   $N_{K^*_J}=  \sqrt{\lambda} 
q^2\beta_l /(96 \pi^3m_{B_s}^3)$. 
The  functions $A_{i}$ can be decomposed into different partial-waves  
\begin{eqnarray}
 A_{0/t }(q^2, m_{K\pi}^2,\theta_K)&=& \sum_{J=0,1,2...}  A^J_{0/t }(q^2,
 m_{K\pi}^2)Y_{J}^0(\theta_K,0),\nonumber\\
 A_{||/\perp }(q^2, m_{K\pi}^2,\theta_K)&=& \sum_{J=0,1,2...}  A^J_{||/
 \perp }(q^2, m_{K\pi}^2)Y_{J}^{-1}(\theta,0),\nonumber \\
% A^J_{0/t }(q^2, m_{K\pi}^2)&=&   \sqrt{ N_{K_J^*}} {\cal M}_B(K^*_J, 0/t )(q^2)
% L_{K^*_J}(m_{K\pi}^2) \equiv | A^J_{ 0/t }| 
% e^{i\delta^J_{{ 0/t }}},\nonumber\\
 A^J_{ i}(q^2, m_{K\pi}^2)&=&   \sqrt{ N_{K_J^*}} {\cal M}_B(K^*_J,  i)(q^2)
 L_{K^*_J}(m_{K\pi}^2)\equiv | A^J_{i}| e^{i\delta^J_{{ i}}}~. 
\end{eqnarray}
 Here,  the subscript $t$ denotes the time-like component of a 
virtual vector/axial-vector meson that decays into the lepton pair.  
$L_{K^*_J}(m_{K\pi})$ is the   lineshape,  and for
the P-wave resonance $K^*(892)$,  we use the Breit-Wigner distribution:
\begin{eqnarray}
 L_{K^*}(m_{K\pi}^2)= \sqrt{ \frac{ m_{K^*} \Gamma_{K^*\to K\pi}  }{\pi} }
 \frac{1}{ m_{K\pi}^2 -m_{K^*}^2+ i m_{K^*}\Gamma_{K^*}}~.
\end{eqnarray}  
Considering the  {momentum dependence
of the $K^*$ decay},  we have the running width 
\begin{eqnarray}
 \Gamma_{K^*} (m_{K\pi}^2) =  \Gamma_{K^*}^0 \left(\frac{ |\vec q\,|}{ |
 \vec q_0|}\right)^3   \frac{m_{K^*}}{m_{K\pi}}   
 \frac{1+ (R|\vec q_0|)^2}{1+ (R|\vec q\,|)^2},\label{eq:KstarLineShape}
\end{eqnarray}
and the Blatt-Weisskopf parameter $R=(2.1\pm 0.5\pm 0.5) {\rm
GeV}^{-1}$~\cite{delAmoSanchez:2010fd}. 
In the following, we will suppress the dependence on the $q^2, m_{K\pi}$, 
and  the polar angle $\theta_K$ for simplicity.

The spin-0 final state has only one polarisation state and the amplitudes are
\begin{eqnarray}
 i{\cal M}_B(K^*_0,0)&=& N_1 i\Bigg[   \frac{\sqrt {\lambda}}{\sqrt{ q^2}} F_1(q^2)  \Bigg],\;\;\;
 i{\cal M}_B(K^*_0,t)=N_1 i\Bigg[ \frac{m_{B_s}^2-m_{K^*_0}^2}{\sqrt {q^2}} F_0(q^2) \Bigg],
\end{eqnarray}
with  $N_1=  {iG_F}V_{ub}/{\sqrt 2}$.
In the case of strange mesons with spin $J\ge1$, the $K^-\pi^+$ system can be 
either longitudinally or transversely polarised and thus we have the following form: 
\begin{eqnarray}
 i{\cal M}_B(K^*_J,0)&=&\frac{ \alpha_L^J N_1  i}{2m_{K^*_J}\sqrt {q^2}}\left[   (m_{B_s}^2-m_{K^*_J}^2-q^2)(m_{B_s}+m_{K^*_J})A_1
 -\frac{\lambda}{m_{B_s}+m_{K^*_J}}A_2\right], \nonumber\\
 i{\cal M}_B({K^*_J},\pm)
 &=& \beta_T^J N_1  i \left[  (m_{B_s}+m_{K^*_J})A_1\mp \frac{\sqrt \lambda}{m_{B_s}+m_{K^*_J}}V \right],\\
 i{\cal M}_B({K^*_J}, L, t)&=&\alpha_L^J i N_{1}  \frac{\sqrt \lambda}{\sqrt {q^2}}A_0~.
\end{eqnarray}
Here, $\alpha_L^J$ and $\beta_T^J$ are products of the   Clebsch-Gordan coefficients
\begin{eqnarray}
 \alpha_L^J &=& C^{J,0}_{1,0;J-1,0} C^{J-1,0}_{1,0; J-2,0} ... C^{2,0}_{1,0;1,0},\;\;\;
 \beta_T^J = C^{J,1}_{1,1;J-1,0} C^{J-1,0}_{1,0; J-2,0} ... C^{2,0}_{1,0;1,0}.
\end{eqnarray}

For the sake of convenience, we can define
\begin{eqnarray}
 i{\cal M}_{B}(K^*_J, \perp/||)&=&\frac{1}{\sqrt 2}[i{\cal M}_B(K^*_J, +) \mp i{\cal M}_B(K^*_J, -)],\nonumber\\
i{\cal M}_B(K^*_J,  \perp) &=& -i\beta_T^J \sqrt{2} N_1\left[ 
 \frac{\sqrt \lambda V}{m_{B_s}+m_{K^*_J}} \right],\nonumber\\
i{\cal M}_B(K^*_J,||)&=& i\beta_T^J\sqrt{2} N_{1} \left[  (m_{B_s}+m_{K^*_J})A_1 \right]~.
\end{eqnarray}

Using the generalised form factor, the  matrix elements for $B_s$ decays 
into the spin-0 non-resonating  $K\pi$  final state  are given as
\begin{eqnarray}
 A_0^0 &=& \sqrt{ N_2} i\frac{1}{m_{K\pi}} \Bigg[   \frac{\sqrt {\lambda}}{\sqrt{ q^2}}{\cal F}_1^{B_s\to K\pi}(m_{K\pi}^2, q^2)  \Bigg]  \equiv \sqrt {N_{K^*_J} } L_{S}(m_{K\pi}^2 ) i \Bigg[   \frac{\sqrt {\lambda}}{\sqrt{ q^2}}{  F}_1^{B_s\to \kappa}(  q^2)  \Bigg] ,\nonumber\\
 A_t^0 &=&\sqrt{ N_2} i \frac{1}{m_{K\pi}}\Bigg[ \frac{m_{B_s}^2-m_{K\pi }^2}{\sqrt {q^2}} {\cal F}_0^{B_s\to K\pi}(m_{K\pi}^2, q^2)\Bigg]  \nonumber\\
 && \equiv \sqrt {N_{K^*_J} } L_{S}(m_{K\pi}^2 )i  \Bigg[ \frac{m_{B_s}^2-m_{K\pi }^2}{\sqrt {q^2}} {  F}_0^{B_s\to \kappa}(m_{K\pi}^2, q^2)\Bigg],\label{eq:S-waveKpi-formula}
\end{eqnarray}  
where we have introduced a lineshape for the S-wave contribution. 
Here $N_2=N_1 N_{K^*_J} \rho_K/(16\pi^2)$, and $\rho_K= \sqrt{[m^2_{K\pi}-(m_K+m_\pi)^2][m^2_{K\pi}-(m_K-m_\pi)^2]}/(m^2_{K\pi})$.  

\smallskip

The   quantities given above lead to the full angular distributions
\begin{eqnarray}
 \frac{d^5\Gamma}{dm_{K\pi}^2dq^2d\cos\theta_K d\cos\theta_l d\phi}
 &=& \frac{3}{8}\Big[I_1(q^2, m_{K\pi}^2, \theta_K)   \nonumber\\
 && +I_2 (q^2, m_{K\pi}^2, \theta_K)  
 \cos(2\theta_l)  \nonumber\\
 && + I_3(q^2, m_{K\pi}^2, \theta_K) \sin^2\theta_l
 \cos(2\phi) \nonumber\\
 &&+I_4(q^2, m_{K\pi}^2, \theta_K)  \sin(2\theta_l)\cos\phi \nonumber\\
 && +I_5 (q^2, m_{K\pi}^2, \theta_K)  \sin(\theta_l) \cos\phi \nonumber\\
 &&+I_6 (q^2, m_{K\pi}^2, \theta_K)  \cos\theta_l \nonumber\\
 && +I_7 (q^2, m_{K\pi}^2, \theta_K) 
 \sin(\theta_l) \sin\phi\nonumber\\
 && +I_8(q^2, m_{K\pi}^2, \theta_K)  \sin(2\theta_l)\sin\phi \nonumber\\
 &&+I_9(q^2, m_{K\pi}^2, \theta_K)  \sin^2\theta_l
 \sin(2\phi)\Big],
\end{eqnarray} 
with the $I_i$ having the form:
\begin{eqnarray}
 I_1   &=& (1+\hat m_l^2) |A_{0}|^2 
 +2 \hat m_l^2  |A_t|^2 +\frac{1}{2} (3+\hat m_l^2)(|A_{\perp}|^2
 +|A_{||}|^2)
 \nonumber\\
 I_2   &=& -\beta_l   |A_{0}|^2+ \frac{1}{2}\beta_l^2 (|A_{\perp}|^2
 +|A_{||}|^2),
 \nonumber\\
  I_3    &=& \beta_l (|A_{\perp}|^2-|A_{||}|^2),\;\;\;\;\;\;\;\;\;\;\;\;\;\;\;\;\;\;\;\;\;
 I_4    =  2 \beta_l  [{\rm Re}(A_{0}A_{||}^*)],\nonumber\\
 I_5   &=&4  [{\rm Re}(A_{0}A_{\perp}^*) -\hat m_l^2 {\rm Re}(A_{t}A_{||}^*) ],\;\;\;
 I_6     =  4  [{\rm Re}(A_{||}A^*_{\perp})+ \hat m_l^2 {\rm Re}(A_{t}A^*_{0})],\nonumber\\
 I_7   &=& 4[{\rm Im}(A_{0}A^*_{||})-\hat m_l^2 {\rm Im}(A_{t}A^*_{\perp})],\;\;\;
 I_8   =  2 \beta_l   [{\rm Im}(A_{0}A^*_{\perp})],\nonumber\\
 I_9  &=&2\beta_l   [{\rm Im}(A_{\perp }A^*_{||})]~. 
\label{eq:angularCoefficients}
\end{eqnarray}
 For the general expressions of these functions, we refer the reader to 
the appendix, and also Ref.~\cite{Lee:1992ih,Kopp:1990yx,Ananthanarayan:2005us}.  
In the following, we shall only consider the S-wave and P-wave contributions and 
thus the above general expressions are reduced to:
\begin{eqnarray}
 I_1  &=& 
   \frac{1}{4\pi} \left[(1+\hat m_l^2) |A^0_{0}|^2  
 +2 \hat m_l^2  |A_t^0|^2\right] + \frac{3}{4\pi} \cos^2\theta_K \left[(1+\hat m_l^2) |A^1_{0}|^2  
 +2 \hat m_l^2  |A_t^1|^2\right]   \nonumber\\
 &&+ \frac{2\sqrt3 \cos\theta_K}{4\pi} \left[  (1+\hat m_l^2) {\rm Re}[A^0_{0} A^{1*}_{0}]   + 2\hat m_l^2 {\rm Re}[A^0_{t} A^{1*}_{t}] \right]
  \nonumber\\
 &+&    \frac{3+\hat m_l^2}{2}  \frac{3}{8\pi}  \sin^2\theta_K    [|A^1_{\perp}|^2+|A^1_{||}|^2 ],\nonumber\\
 %\end{eqnarray} 
 %%%%
 %\begin{eqnarray}
 I_2   &=& -\beta_l     \bigg\{ \frac{1}{4\pi} |A^0_{0}|^2 + \frac{3}{4\pi}\cos^2\theta_K    |A^1_{0}|^2   + \frac{2\sqrt3 \cos\theta_K}{4\pi}  
 {\rm Re}[A^0_{0} A^{1*}_{0}]   \bigg\} \nonumber\\
% \end{eqnarray} 
 %%%%
% \begin{eqnarray}
 %%%%%%%%%
  &+&
 \frac{1}{2}\beta_l    \frac{3}{8\pi} \sin^2\theta_K    (|A^1_{\perp}|^2+|A^1_{||}|^2),
 \nonumber\\
% \end{eqnarray} 
 %%%%
% \begin{eqnarray}
 %%%%%%%%%
 I_3&=&
 \beta_l    \frac{3}{8\pi} \sin^2\theta_K    (|A^1_{\perp}|^2-|A^1_{||}|^2),
 \nonumber\\
% \end{eqnarray} 
 %%%%
% \begin{eqnarray}
 %%%%%%%
 I_4
  &=&2   \beta_l     \left[ \frac{\sqrt 3 \sin\theta_K}{4\sqrt 2\pi} {\rm Re}[A^0_{0}A^{1*}_{||} ] + \frac{3\sin\theta_K\cos\theta_K }{4\sqrt 2\pi}   {\rm Re}[A^1_{0}A^{1*}_{||} ]  \right],
  \nonumber
 \end{eqnarray} 
 %%%%
 \begin{eqnarray}  %%%%%%%%%%%
 I_5
  &=&4\bigg\{\frac{\sqrt 3 \sin\theta_K}{4\sqrt 2\pi}  ({\rm Re}[A^0_{0}A^{1*}_{\perp } ] -\hat m_l^2  {\rm Re}[A^0_{t}A^{1*}_{||} ] )    \nonumber\\
  &&  + \frac{  3 \sin\theta_K\cos\theta_K }{4\sqrt 2\pi} ( {\rm Re}[A^1_{0}A^{1*}_{\perp } ] -\hat m_l^2     {\rm Re}[A^1_{t}A^{1*}_{||} ]) \bigg\} ,
   \nonumber\\
% \end{eqnarray} 
 %%%%
% \begin{eqnarray}
  %%%%%%%%%
 I_6&=& 4\bigg\{ \frac{3}{8\pi} \sin^2\theta_K   {\rm Re}[ A^1_{||}A^{1*}_{\perp} ]  +  \hat m_l^2 \frac{1}{4\pi} {\rm Re}[A_t^0 A_{0}^{0*}] + \hat m_l^2 \frac{3}{4\pi} \cos^2\theta_K {\rm Re}[A_t^1 A_{0}^{1*}] \bigg\}
 \nonumber\\
% \end{eqnarray} 
 %%%%
% \begin{eqnarray}
  I_7
 &=& 4\bigg\{\frac{\sqrt 3}{4\sqrt 2\pi} \sin\theta_K   ({\rm Im}[A^0_{0}A^{1*}_{||}] - \hat m_l^2 {\rm Im}[ A_t^0 A_\perp^{1*}] )   \nonumber\\
 && +    \frac{3}{4\sqrt 2\pi} \sin\theta_K \cos\theta_K  ({\rm Im}[A^1_{0}A^{1*}_{||}] - \hat m_l^2 {\rm Im}[ A_t^1 A_\perp^{1*}] )\bigg\}  \nonumber\\
% \end{eqnarray} 
 %%%%
% \begin{eqnarray}
 %%%%%%%
 I_8 &=&
 2 \beta_l  \bigg\{\frac{\sqrt 3}{4\sqrt 2\pi} \sin\theta_K    {\rm Im} [A_0^0 A_\perp^{1*}]+ \frac{  3}{4\sqrt 2\pi} \sin\theta_K \cos\theta_K   {\rm Im} [A_0^1 A_\perp^{1*}]\bigg\},
 \nonumber\\
% \end{eqnarray} 
 %%%%
% \begin{eqnarray}
 %%%%%%%%
 I_9
 &=&2 \beta_l  \frac{3}{8\pi} \sin^2\theta_K {\rm Im}[A_\perp^1 A_{||}^{1*} ]~.
 \label{eq:simplified_angularCoefficients}
\end{eqnarray} 
 One difference compared to the $B\to K\pi l^+l^-$ distributions~\cite{Lu:2011jm,Doring:2013wka}, where the leptons have the same mass in the above coefficients, are:  in $I_{5,6,7}$,  the time component is present and interferes with the transverse polarisation. These are finite  lepton mass corrections. Moreover, 
since the phase in the P-wave contributions arise from the line-shape which 
is the same for different polarisations, the $I_9$ term  and the second line   
in the $I_7$ are zero here. 

Measurements of  the $B_s\to K\pi \ell\bar\nu$ and $B\to \pi\pi\ell\bar\nu$  processes can test the $\Delta I=1/2$ rule~\cite{kl4,Ananthanarayan:2005us} since they  are both induced by the $b\to u\ell\bar\nu$. However,  channels with a neutral $\pi^0$ in the final state will  request a high statistics to be accumulated in future experimental facilities. 

These decay modes can also provide a probe for the T violation~\cite{kl4,Ananthanarayan:2005us}.  
As we have shown in the above,  if only S-wave and P-wave are considered, the $I_9$ term  and the second line   
in the $I_7$ are zero.  These coefficients  can be nonzero  either due to the violation of T-invariance or  higher partial-wave contribution.  Furthermore, comparing the distributions for the $B^-$ decay and its CP conjugation mode can examine  the T-invariance as well, however it is difficult  to generalize to  the neutral $B$ decays, since the $B^0$ and $B_s^0$  mix with their CP partner, respectively. 

\subsection{Differential and integrated decay widths}

The starting point for a detailed analysis is to obtain the double-differential 
distribution  $d^2\Gamma/dq^2/dm_{K\pi}^2$  after performing an integration 
over all the angles 
\begin{eqnarray}
\frac{d^2\Gamma}{dq^2 dm_{K\pi}^2} &=&   \big(1+\frac{\hat m_l^2}{2}\big)( |A_{0}^0|^2 +  |A_{0}^1|^2  + |A_{||}^1|^2 +  |A_{\perp}^1|^2 )   + \frac{3}{2} \hat m_l^2 (|A_{t}^1|^2 + |A_{t}^0|^2 ),
\end{eqnarray}
where apparently in the massless limit for the involved lepton,  
the total normalization for the angular distributions changes to the sum 
of the S-wave and P-wave amplitudes 
\begin{eqnarray}
\frac{d^2\Gamma}{dq^2 dm_{K\pi}^2} =   |A_{0}^0|^2 +  |A_{0}^1|^2  + |A_{||}^1|^2 +  |A_{\perp}^1|^2. 
\end{eqnarray}

To match the kinematics constraints  implemented in the experimental measurements, 
one may explore the generic observable  with $m_{K\pi}^2$ integrated out: 
\begin{eqnarray}
 \langle O\rangle = \int_{(m_{K^*}-\delta_m)^2}^{(m_{K^*}+\delta_m)^2} dm^2_{K\pi} \frac{dO}{dm_{K\pi}^2}~. 
\end{eqnarray}
Following the recent  LHCb measurements on  $B\to K^*(892)l^+l^-$~\cite{Aaij:2013iag}, 
we use the following choice  in our study of $B_s\to K\pi \ell\bar\nu$: 
\begin{eqnarray}
 \delta_m =100 {\rm MeV}.
\end{eqnarray}
In the narrow width-limit for the   P-wave contributions, the integration of 
the lineshape gives
\begin{eqnarray}
 \int dm^2_{K\pi}|L_{K^*}(m^2_{K\pi})|^2 = {\cal B}(K^{*+}\to K^0\pi^+)=\frac{2}{3}~. 
\end{eqnarray}
However with the explicit form given in Eq.\eqref{eq:KstarLineShape}, we find  
that the integration 
\begin{eqnarray}
 \int_{(m_{K^*}-\delta_m)^2}^{(m_{K^*}+\delta_m)^2} dm^2_{K\pi}|L_{K^*}(m^2_{K\pi})|^2 =0.56,  
\end{eqnarray}
is below the expected value. 
This  mismatch clearly indicates that one should be cautious to identify the   
experimental signal  with theoretical results based on the purely P-wave contributions.  
On the other hand,  the integrated  S-wave lineshape in this region is
\begin{eqnarray} 
\int_{(m_{K^*}-\delta_m)^2}^{(m_{K^*}+\delta_m)^2} dm^2_{K\pi}|L_{S}(m^2_{K\pi})|^2 =0.17,
\end{eqnarray}
which is at the same order and can not be neglected. Future experimental  
measurements should take this into account.

 Furthermore, one may   explore the $q^2$-dependent ratio
\begin{eqnarray}
\label{eq20:DDRRDst2}
R_{K\pi}^{\tau/\mu}(q^2) &=&\frac{\langle d\Gamma(\overline B_s\to K\pi \tau\bar\nu)/dq^2 \rangle }{\langle d\Gamma(\overline  B_s\to K\pi l\bar\nu)/dq^2\rangle }\,, \label{eq:RKpitauOverMu}
\end{eqnarray}
where $l$ denotes the light lepton $(e, \mu)$.  This ratio will be less 
sensitive to S-wave contributions.

\begin{figure}[t]
\begin{center}
\includegraphics[scale=0.5]{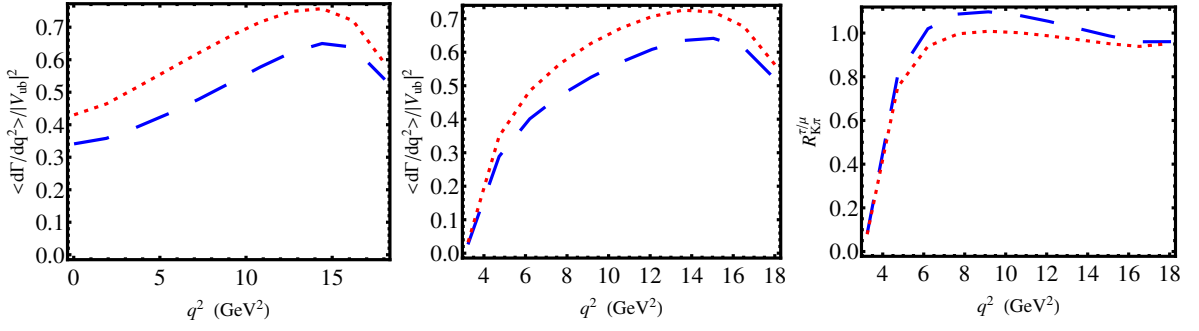} 
\caption{ Differential decay widths  (in units of ${\rm ps}^{-1}  {\rm GeV}^{-2} $)  
for  $\overline B_s\to K^0\pi^+\ell \bar\nu_{\ell}$ with $\ell= \mu,e$ in the first panel   
and $\ell=\tau$ in the second  panel. The $q^2$-dependent ratio $R_{K\pi}^{\tau/\mu}$ as 
defined in Eq.~\eqref{eq:RKpitauOverMu} is given in the last panel. The dashed and 
dotted curves are obtrained using the LQCD and LCSR results for the 
$B_s\to K^*$ form factors.  } \label{fig:diffdwBsKstar}
\end{center}
\end{figure}

Differential decay widths for  $B_s\to K\pi\ell \bar\nu_{\ell}$ are  
given in Fig.~\ref{fig:diffdwBsKstar},  with $\ell= \mu,e$ 
and $\ell=\tau$ in the first two panels respectively. The $q^2$-dependent ratio $R_{K\pi}^{\tau/\mu}$ 
as defined in Eq.~\eqref{eq:RKpitauOverMu} is given in the last  panel.

Integrating over $q^2$, one has the partial width 
\begin{eqnarray}
 \Delta \zeta_{K\pi}^{\ell} (q_{l}^2, q_{u}^2) 
= \frac{1}{|V_{ub}|^2}  \int_{q_l^2}^{q_u^2} dq^2 \left\langle \frac{d\Gamma}{dq^2}\right\rangle~. 
\end{eqnarray}

With the form factors from LQCD~\cite{Horgan:2013hoa} and the LCSR~\cite{Ball:2004rg}, 
we give numerical results for $ \Delta \zeta_{K^0\pi^+}^{\ell} (q_{l}^2,q_u^2)$ with 
different sets of $q_l$ and $q_u$  in  Tab.~\ref{tab:integratedWidthBstoKpi}. 
The S-wave contribution is calculated using the $K\pi$ scalar form factors 
and the $B_s\to \kappa$ transitions. Its contribution to the integrated 
partial widths ranges from $10\%$ to approximately $20\%$. Since the 
S-wave contains the factor $1/\sqrt{q^2}$,  as shown in Eq.\eqref{eq:S-waveKpi-formula}, 
its contribution  decreases with the increasing  $q^2$. 
 
%%%%%%%%%%%%%%%%%%%%%%%%%%%%%%%%%%%%%%%%%%%%%%%%%%%%%%%%%%%%%%%%%%%%%%%%%%%%%%%%
%%%%%%%%%%%%%%%%%%%%%%%%%%%%%%%%%%%%%%%%%%%%%%%%%%%%%%%%%%%%%%%%%%%%%%%%%%%%%%%% 

%%%%%%%%%%%%%%%%%%%%%%%%%%%%%%%%%%%%%%%%%%%%%%%%%%%%%%%%%%%%%%%%%%%%%%%%%%%%%%%% 
 %\input{tableBsKstar.tex}
 \begin{table}[t]
\begin{center}
 \caption{Integrated decay widths for
 $\overline B_s\to K^0 \pi^+  \ell\bar  \nu$:
 $\Delta \zeta_{K\pi}^{\ell} (q_l^2, q_u^2)$.
Results are given in units of $ps^{-1}$.
The S-wave contributions are obtained 
with form factors calculated in the PQCD approach.}
\label{tab:integratedWidthBstoKpi}
\begin{tabular}{cc|cccccccc}
\hline\hline
  & & S-Wave & P-Wave & Total &$f_S(\%)$  \\\hline
$\Delta \zeta_{K\pi}^\mu(0, 4)$
 & LQCD  & $0.278\pm 0.151 $ & $ 1.19\pm 0.151 $&$ 1.47\pm 0.299 $&$ 18.9 $ 
\\
 & LCSR  & $0.278\pm 0.151 $ & $ 1.61\pm 0.172 $&$ 1.89\pm 0.321 $&$ 14.7 $ 
\\
 $\Delta \zeta_{K\pi}^\mu(4, 8)$ 
 & LQCD  & $0.276\pm 0.149 $ & $ 1.58\pm 0.221 $&$ 1.85\pm 0.376 $&$ 14.9 $ 
\\
 & LCSR  & $0.276\pm 0.149 $ & $ 2.13\pm 0.296 $&$ 2.41\pm 0.441 $&$ 11.5 $ 
\\
$\Delta \zeta_{K\pi}^\mu(8, 12)$
 & LQCD  & $0.254\pm 0.137 $ & $ 2.03\pm 0.305 $&$ 2.29\pm 0.436 $&$ 11.1 $ 
\\
 & LCSR  & $0.254\pm 0.137 $ & $ 2.6\pm 0.421 $&$ 2.86\pm 0.552 $&$ 8.88 $ 
\\
$\Delta \zeta_{K\pi}^\mu(12, 16)$
 & LQCD  & $0.202\pm 0.109 $ & $ 2.43\pm 0.347 $&$ 2.63\pm 0.458 $&$ 7.68 $ 
\\
 & LCSR  & $0.202\pm 0.109 $ & $ 2.88\pm 0.505 $&$ 3.08\pm 0.616 $&$ 6.56 $ 
\\
$\Delta \zeta_{K\pi}^\mu(0, 16)$
 & LQCD  & $1.01\pm 0.546 $ & $ 7.22\pm 1.03 $&$ 8.23\pm 1.58 $&$ 12.3 $ 
\\
 & LCSR  & $1.01\pm 0.546 $ & $ 9.23\pm 1.38 $&$ 10.2\pm 1.97 $&$ 9.9 $ 
\\
\hline
 $\Delta \zeta_{K\pi}^\tau(m_\tau^2, 8)$
 & LQCD  & $0.281\pm 0.152 $ & $ 1.36\pm 0.239 $&$ 1.64\pm 0.392 $&$ 17.1 $ 
\\
 & LCSR  & $0.281\pm 0.152 $ & $ 1.7\pm 0.253 $&$ 1.98\pm 0.407 $&$ 14.2 $ 
\\
$\Delta \zeta_{K\pi}^\tau(8, 12)$
 & LQCD  & $0.299\pm 0.162 $ & $ 1.97\pm 0.319 $&$ 2.27\pm 0.48 $&$ 13.2 $ 
\\
 & LCSR  & $0.299\pm 0.162 $ & $ 2.38\pm 0.399 $&$ 2.68\pm 0.56 $&$ 11.2 $ 
\\
 $\Delta \zeta_{K\pi}^\tau(12, 16)$
 & LQCD  & $0.552\pm 0.299 $ & $ 4.32\pm 0.666 $&$ 4.88\pm 0.958 $&$ 11.3 $ 
\\
 & LCSR  & $0.552\pm 0.299 $ & $ 5.09\pm 0.873 $&$ 5.64\pm 1.17 $&$ 9.79 $ 
\\
 $\Delta \zeta_{K\pi}^\tau(m_\tau^2, 16)$
 & LQCD  & $0.834\pm 0.45 $ & $ 5.69\pm 0.895 $&$ 6.52\pm 1.35 $&$ 12.8 $ 
\\
 & LCSR  & $0.834\pm 0.45 $ & $ 6.78\pm 1.14 $&$ 7.62\pm 1.58 $&$ 10.9 $ 
\\
\hline
 \hline
\end{tabular} 
\end{center}
\end{table}

 %%%%%%%%%%%%%%%%%%%%%%%%%%%%%%%%%%%%%%%%%%%%%%%%%%%%%%%%%%%%%%%%%%%%%%%%%%%%%%%%

 \subsection{Distribution in {\boldmath$\theta_K$} }

%%%%%%%%%%%%%%%%%%%%%%%%%%%%%%%%%%%%%%%%%%%%%%%%%%%%%%%%%%%%%%%%%%%%%%%%%%%%%%%%
\begin{figure}[t]
\begin{center}
\includegraphics[scale=0.475]{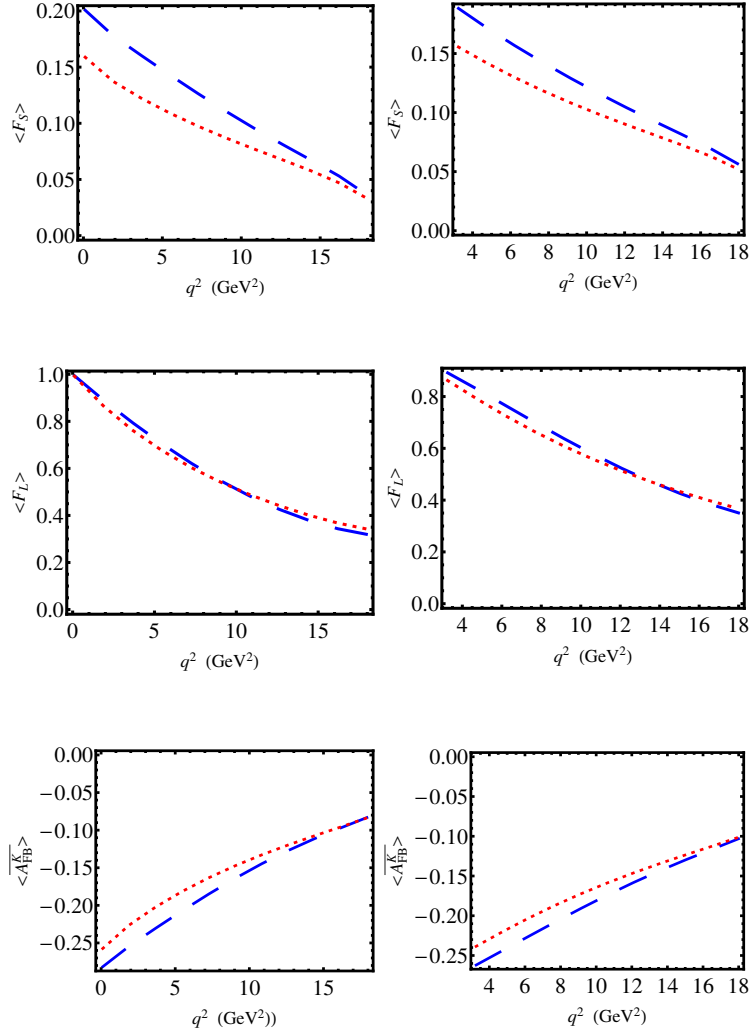} 
\caption{Same as Fig.~\ref{fig:diffdwBsKstar} but for the S-wave contribution (a,b) 
and the longitudinal  polarisations in the P-wave contribution (c,d)  to 
the $B_s\to K\pi \ell\bar\nu_{\ell}$, and the forward-backward asymmetry 
$\overline {A_{FB}^{K}}$ (e,f). Notice that for $\overline {A_{FB}^{K}}$, there 
is a sign ambiguity arising from the use of the Watson theorem. The left 
panels  are for the light lepton $e,\mu$ while the right panels   are for the $\tau$ lepton.   } \label{fig:FSBsKstar}
\end{center}
\end{figure}
%%%%%%%%%%%%%%%%%%%%%%%%%%%%%%%%%%%%%%%%%%%%%%%%%%%%%%%%%%%%%%%%%%%%%%%%%%%%%%%% 

We explore the distribution in $\theta_K$: 
\begin{eqnarray}
\frac{d^3\Gamma} { dq^2 dm_{K\pi}^2 d\cos\theta_K}  
%&=&  \frac{\pi}{2} (3I_1-I_2) \nonumber\\
&=& \frac{1}{8} \bigg\{ (4+2\hat m_l^2) |A_0^0|^2 + 6\hat m_l^2 |A_t^0|^2 \nonumber\\
&&  + \sqrt{3} (8+4\hat m_l^2) \cos\theta_K {\rm Re}[A_0^0 A_{0}^{1*}] + 12\sqrt{3} \hat m_l^2 \cos\theta_K  {\rm Re}[A_t^0 A_{t}^{1*}] \nonumber\\
&&+ (12+ 6\hat m_l^2) |A_0^1|^2 \cos^2\theta_K + 18\hat m_l^2 \cos^2\theta_K |A_t^1|^2   \nonumber\\
&&  + (6+3\hat m_l^2) \sin^2\theta_K  (|A_\perp^1|^2 + |A_{||}^1|^2)  \bigg\}~.
\label{eq:theta_K_distribution} 
\end{eqnarray}

Compared to the distribution with only $B_s\to K^*(\to K\pi)\ell\bar\nu$,  
the first two lines   of Eq.~\eqref{eq:theta_K_distribution} are new: the first one is
the  S-wave $K\pi$ contribution,  while the second line  corresponds to  the
interference of   S-wave and P-wave.  Based on this interference,  one can  
define a forward-backward  asymmetry for the involved  hadron, 
\begin{eqnarray}
 A_{FB}^{K} &\equiv &   \bigg[\int_{0}^1 
 - \int_{-1}^0\bigg] d\cos\theta_K \frac{d^3\Gamma}{ dq^2 dm_{K\pi}^2
  d\cos\theta_K}  \nonumber\\
 &=&  \frac{\sqrt{3}}{2}  (2+ \hat m_l^2)  {\rm Re}[A_0^0 A_{0}^{1*}] + \frac{3\sqrt{3}}{2}  \hat m_l^2   {\rm Re}[A_t^0 A_{t}^{1*}]~.
\end{eqnarray}

%%%%%%%%%%%%%%%%%%%%%%%%%%%%%%%%%%%%%%%%%%%%%%%%%%%%%%%%%%%%%%%%%%%%%%%%%%%%%%%%
%%\begin{figure}
%%\begin{center}
%%\includegraphics[scale=0.6]{ASBsToKstar.eps} 
%%\caption{The asymmetry $\overline {\cal A}_{S} $   in $B_s\to K\pi \ell\bar\nu_{\ell}$} \label{fig:ASBsKstar}
%%\end{center}
%%\end{figure}
%%%%%%%%%%%%%%%%%%%%%%%%%%%%%%%%%%%%%%%%%%%%%%%%%%%%%%%%%%%%%%%%%%%%%%%%%%%%%%%% 

We define the polarisation fraction  at a given value of $q^2$ and $m_{K\pi}^2$: 
\begin{eqnarray}
 {\cal F}_{S} (q^2, m_{K\pi}^2)  = \frac{ (1+\hat m_l^2/2) |A_0^0|^2 + 3/2 \hat m_l^2 |A_t^0|^2 } {  {d^2\Gamma}/({dq^2 dm_{K\pi}^2})} ,\nonumber\\
 {\cal F}_{P } (q^2, m_{K\pi}^2)  = \frac{  (1+\hat m_l^2/2)(|A_{0}^1|^2  + |A_{||}^1|^2 +  |A_{\perp}^1|^2 ) + 3/2 |A^1_t|^2 } { {d^2\Gamma}/({dq^2 dm_{K\pi}^2})} ,
\end{eqnarray}
and also 
\begin{eqnarray}
 F_L (q^2, m_{K\pi}^2)  = \frac{   (1+\hat m_l^2/2)(|A_{0}^1(q^2, m_{K\pi}^2)|^2  + 3/2 |A^1_t|^2  } {   (1+\hat m_l^2/2)(|A_{0}^1|^2  + |A_{||}^1|^2 +  |A_{\perp}^1|^2 ) + 3/2 |A^1_t|^2 }.
\end{eqnarray}
By definition, ${\cal F}_S+{\cal F}_P=1$.

In Fig.~\ref{fig:FSBsKstar}.
we give our results for the S-wave fraction $\langle  F_S\rangle$, longitudinal polarisation fraction $\langle  F_L\rangle$  in the
P-wave contributions and the asymmetry $\langle  
\overline {A_{FB}^{K}} \rangle$. The curves in the left panels  are 
for the light lepton $e,\mu$ while the right three panels  are for the $\tau$ lepton. 
These observables and the following ones  are defined via the integration 
over $m_{K\pi}^2$, for instance 
\begin{eqnarray}
 \langle  F_{S} (q^2) \rangle = \frac{\int dm_{K\pi}^2 [ (1+\hat m_l^2/2) |A_0^0|^2 + 3/2 \hat m_l^2 |A_t^0|^2]  } {\int dm_{K\pi}^2  {d^2\Gamma}/({dq^2 dm_{K\pi}^2})} ~,
\end{eqnarray}
and likewise for the others.

\subsection{Distribution in {\boldmath$\theta_l$} and forward-backward asymmetry}

%%%%%%%%%%%%%%%%%%%%%%%%%%%%%%%%%%%%%%%%%%%%%%%%%%%%%%%%%%%%%%%%%%%%%%%%%%%%%%%%
\begin{figure}[bth]
\begin{center}
\includegraphics[scale=0.6]{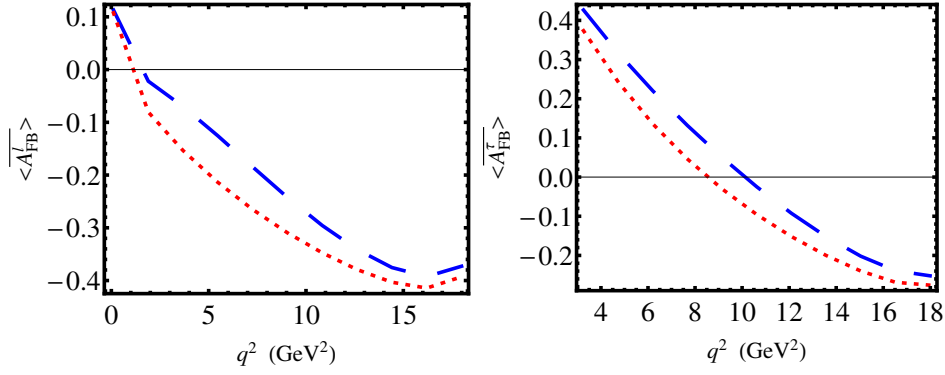} 
\caption{Same as Fig.~\ref{fig:diffdwBsKstar} but for the asymmetry 
$\overline {{\cal A}_{FB}^l} $ in  $B_s\to K\pi \ell\bar\nu_{\ell}$} \label{fig:AFBBsKstar}
\end{center}
\end{figure}
%%%%%%%%%%%%%%%%%%%%%%%%%%%%%%%%%%%%%%%%%%%%%%%%%%%%%%%%%%%%%%%%%%%%%%%%%%%%%%%% 

After integrating over  $\theta_K$ and $\phi$, we have the distribution: 
\begin{eqnarray}
\frac{d^3\Gamma} { dq^2 dm_{K\pi}^2 d\cos\theta_l}  &=& \frac{3\pi}{4} \int d\cos\theta_K (I_1 +I_2\cos(2\theta_l) +I_6 \cos\theta_l )\nonumber\\
&=& \frac{3}{4} \hat m_l^2 ( |A_{t}^0|^2 +|A_{t}^1|^2 ) \nonumber\\
&& + \frac{3}{2} \cos\theta_l  ( {\rm Re}[A_{||}^1 A_{\perp}^{1*}] + \hat m_l^2  {\rm Re}[A_{t}^0 A_{0}^{0*}+A_{t}^1 A_{0}^{1*}] ) 
\nonumber\\
&& +\frac{3}{4} [1 -(1-\hat m_l^2)\cos^2\theta_l] (|A_{0}^0|^2 +|A_{0}^1|^2 ) \nonumber\\
&& + \frac{3}{8} [ (1+\hat m_l^2) + (1-\hat m_l^2) \cos^2\theta_l ]  (|A_{||}^1|^2 +|A_{\perp}^1|^2 ).
\end{eqnarray} 
The forward-backward asymmetry is defined as
\begin{eqnarray}
 A_{FB}^{l}  &\equiv &   \bigg[\int_{0}^1 
 - \int_{-1}^0\bigg] d\cos\theta_l \frac{d^3\Gamma}{ dq^2 dm_{K\pi}^2
  d\cos\theta_l }  \nonumber\\
 &=&  \frac{3}{2}   ( {\rm Re}[A_{||}^1 A_{\perp}^{1*}] + \hat m_l^2  {\rm Re}[A_{t}^0 A_{0}^{0*}+A_{t}^1 A_{0}^{1*}] ) ,
\end{eqnarray}
and the results  for $\overline {{\cal A}_{FB}^l} $  are given in Fig.~\ref{fig:AFBBsKstar}. 
The $A_{||}^1$ and $A_{\perp}^1$ have different signs, and thus the $\overline {{\cal A}_{FB}^l} $ becomes negative when the $\hat m_l^2$ corrections become less important in the large $q^2$ region.

\subsection{Distribution in the azimuthal  angle {\boldmath $\phi$} } 

%\begin{eqnarray}
% {\cal A}_{\rm Im} &=& \frac{{\rm Im}[A_\perp^1 A_{||}^{1*} ]} {  |A_{0}^1(q^2, m_{K\pi}^2)|^2  + |A_{||}^1(q^2, m_{K\pi}^2)|^2 +  |A_{\perp}^1(q^2, m_{K\pi}^2)|^2},\nonumber\\ 
% {\cal A}_{T}^{2}  &=& \frac{ |A_{\perp}^1(q^2, m_{K\pi}^2)|^2- |A_{||}^1(q^2, m_{K\pi}^2)|^2 } {  |A_{||}^1(q^2, m_{K\pi}^2)|^2 +  |A_{\perp}^1(q^2, m_{K\pi}^2)|^2} ,
%\nonumber\\  {\cal A}_{T}^{1} &=& \frac{ -2{\rm Re}[A_{||}^1 A_{\perp}^{1*}]} {  |A_{||}^1(q^2, m_{K\pi}^2)|^2 +  |A_{\perp}^1(q^2, m_{K\pi}^2)|^2}
%\end{eqnarray}

%%%%%%%%%%%%%%%%%%%%%%%%%%%%%%%%%%%%%%%%%%%%%%%%%%%%%%%%%%%%%%%%%%%%%%%%%%%%%%%%
\begin{figure}
\begin{center}
\includegraphics[scale=0.5]{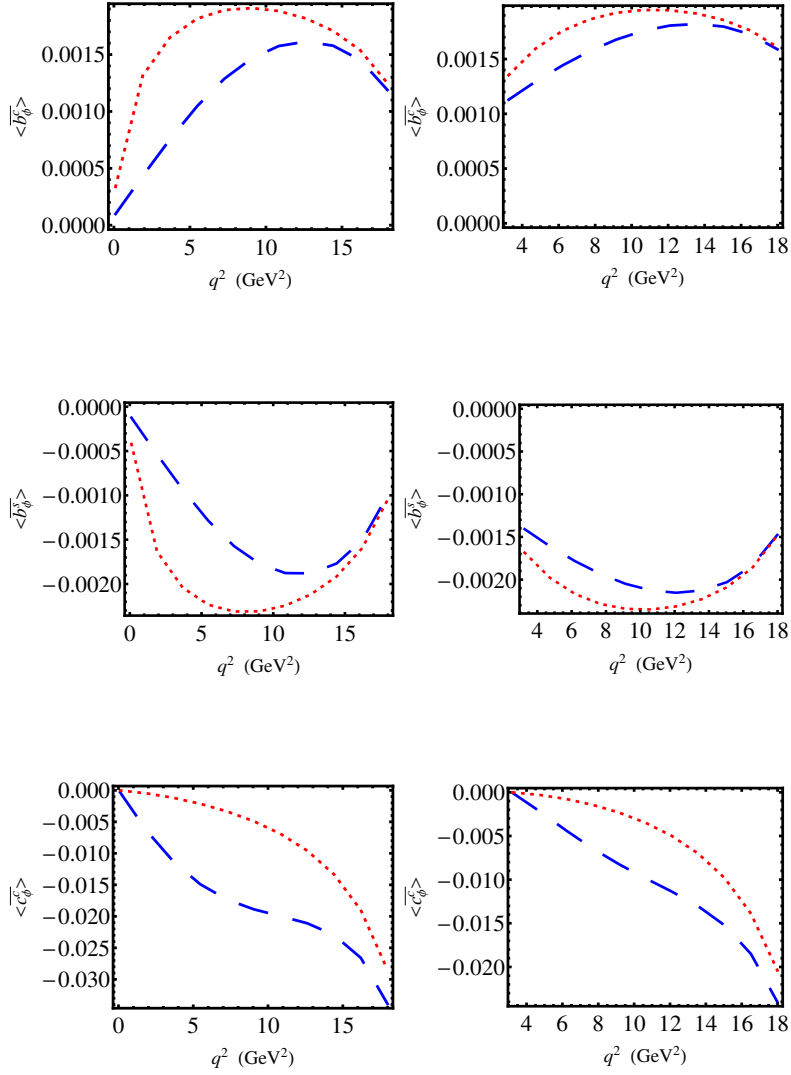} 
\caption{Same as Fig.~\ref{fig:diffdwBsKstar} but for the normalised 
coefficients in the $\phi$ distributions of  $\overline B_s\to K^0\pi^+ \ell\bar\nu_{\ell}$} 
\label{fig:bcphiBsKstar}
\end{center}
\end{figure}
%%%%%%%%%%%%%%%%%%%%%%%%%%%%%%%%%%%%%%%%%%%%%%%%%%%%%%%%%%%%%%%%%%%%%%%%%%%%%%%% 

The angular distribution in $\phi$ is derived  as
\begin{eqnarray}
\frac{d^3\Gamma} { dq^2 dm_{K\pi}^2 d\phi}  &=& a_\phi +b_\phi^c \cos\phi + b_\phi^s \sin\phi + c_\phi^c \cos(2\phi) + c_\phi^s \sin(2\phi), 
\end{eqnarray}
with 
\begin{eqnarray}
 a_\phi &=& \frac{1}{2\pi} \frac{d^2\Gamma} { dq^2 dm_{K\pi}^2}  ,\nonumber\\
 b_\phi^c &=& \frac{3}{16}\pi \int d\cos\theta_K I_5 =  \frac{3\sqrt3}{32\sqrt 2\pi} \big({\rm Re}[A_0^0 A_{\perp}^{1*}]-\hat m_l^2 {\rm Re}[A_t^0 A_{\perp}^{1*}]\big),\nonumber\\
 b_\phi^s &=& \frac{3}{16}\pi \int d\cos\theta_K I_7 =  \frac{3\sqrt3}{32\sqrt 2\pi} \big({\rm Im}[A_0^0 A_{\perp}^{1*}]-\hat m_l^2 {\rm Im}[A_t^0 A_{\perp}^{1*}]\big),\nonumber\\
 c_\phi^c &=& \frac{1}{2}   \int d\cos\theta_K I_3 =  \frac{1}{4\pi} \beta_l (|A_{\perp}^1|^2-|A_{||}^1|^2), \nonumber\\
 c_\phi^s &= &\frac{1}{2}   \int d\cos\theta_K I_9 =  \frac{1}{ 2 \pi } \beta_l {\rm Im}[A_{\perp}^1 A_{||}^{1*}]~.  
\end{eqnarray}
Since the complex phase in the P-wave amplitudes comes from the Breit-Wigner 
lineshape, the coefficient $c_\phi^s$ vanishes. 

Numerical results for the normalised coefficients  using the two sets of form factors   
are shown in Fig.~\ref{fig:bcphiBsKstar}, with the left panels for the light lepton 
and the right ones for the $\tau$ lepton, respectively. The coefficients $ b_\phi^c$ 
and $ b_\phi^s$ contain a very small prefactor, $ {3\sqrt3}/({32\sqrt 2\pi} )\sim 0.037$, 
and thus are numerically tiny   as shown in this figure. The $ c_\phi^c$ is also small 
due to the cancellation between the $|A_{\perp}|^2$ and $|A_{||}|^2$. 

%\begin{eqnarray}
%\frac{d^3\Gamma^{(2)}} { dq^2 dm_{K\pi}^2 d\phi}  &=&  \left[\int_0^1 -  \int_{-1}^0\right] d\cos\theta_K \int_0^1 d\cos\theta_l \frac{d^5\Gamma} { dq^2 dm_{K\pi}^2 d\cos\theta_K d\cos\theta_l d\phi} 
%\end{eqnarray}

%\begin{eqnarray}
%\frac{d^3\Gamma^{(3)}} { dq^2 dm_{K\pi}^2 d\phi}  &=&  \left[\int_0^1 -  \int_{-1}^0\right] d\cos\theta_K \left[\int_0^1 -  \int_{-1}^0\right]  d\cos\theta_l \frac{d^5\Gamma} { dq^2 dm_{K\pi}^2 d\cos\theta_K d\cos\theta_l d\phi} 
%\end{eqnarray}

\subsection{Polarisation of the {\boldmath$\tau$} lepton}

%%%%%%%%%%%%%%%%%%%%%%%%%%%%%%%%%%%%%%%%%%%%%%%%%%%%%%%%%%%%%%%%%%%%%%%%%%%%%%%%
\begin{figure}
\begin{center}
\includegraphics[scale=0.6]{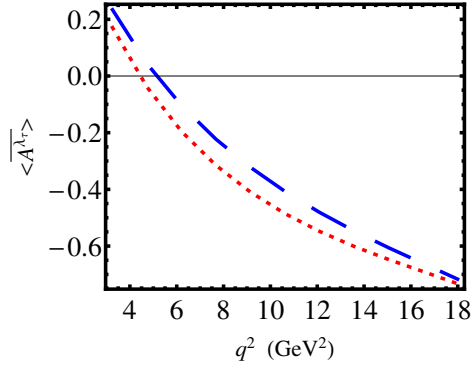} 
\caption{Same as Fig.~\ref{fig:diffdwBsKstar} but for the  polarisation distribution  of $B_s\to K\pi \tau\bar\nu_{\tau}$} \label{fig:AlambdaBBsKstar}
\end{center}
\end{figure}
%%%%%%%%%%%%%%%%%%%%%%%%%%%%%%%%%%%%%%%%%%%%%%%%%%%%%%%%%%%%%%%%%%%%%%%%%%%%%%%% 

We also give the polarised angular distributions as 
\begin{eqnarray}
 \frac{d^5\Gamma(\lambda_\tau)}{dm_{K\pi}^2dq^2d\cos\theta_K d\cos\theta_l d\phi}
 &=& \frac{3}{8}\Big[I_1^{(\lambda_\tau)}   +I_2 ^{(\lambda_\tau)}
 \cos(2\theta_l)   + I_3^{(\lambda_\tau)} \sin^2\theta_l
 \cos(2\phi) \nonumber\\
 &&+I_4^{(\lambda_\tau)}  \sin(2\theta_l)\cos\phi  +I_5^{(\lambda_\tau)} \sin(\theta_l) \cos\phi  +I_6 ^{(\lambda_\tau)}  \cos\theta_l \nonumber\\
 && +I_7^{(\lambda_\tau)}
 \sin(\theta_l) \sin\phi  +I_8^{(\lambda_\tau)}  \sin(2\theta_l)\sin\phi  +I_9^{(\lambda_\tau)} \sin^2\theta_l
 \sin(2\phi)\Big],\nonumber\\
\end{eqnarray}  
with the coefficients
\begin{eqnarray}
 I_1^{(-1/2)}    &=&  |A_{0}|^2  +\frac{3}{2} (|A_{\perp}|^2
 +|A_{||}|^2),
 \nonumber\\
 I_2^{(-1/2)}  &=& -  |A_{0}|^2+ \frac{1}{2}  (|A_{\perp}|^2
 +|A_{||}|^2),
 \nonumber\\
  I_3^{(-1/2)}    &=&   |A_{\perp}|^2-|A_{||}|^2 ,\;\;\;
 I_4^{(-1/2)}    =  2[{\rm Re}(A_{0}A_{||}^*)],\nonumber\\
 I_5^{(-1/2)}  
  &=&4  [{\rm Re}(A_{0}A_{\perp}^*)  ], \;\;\;
 I_6^{(-1/2)}     =  4
  [{\rm Re}(A_{||}A^*_{\perp}) ],\nonumber\\
   I_7^{(-1/2)}   &=& 4[{\rm Im}(A_{0}A^*_{||}) ], \;\;\;
 I_8^{(-1/2)}     =  2   [{\rm Im}(A_{0}A^*_{\perp})],\nonumber\\
 I_9^{(-1/2)}  &=&2   [{\rm Im}(A_{\perp }A^*_{||})].\label{eq:polarisedAngularCoefficients}
\end{eqnarray} 
The coefficients for  $\lambda_\tau=1/2$ are easily obtained by comparing 
Eq.~\eqref{eq:polarisedAngularCoefficients} and Eq.~\eqref{eq:simplified_angularCoefficients}. 

The polarisation fraction for the $\tau$-lepton is defined as
\begin{eqnarray}
 \overline  {A^{\lambda_\tau}}(q^2, m_{K\pi}^2 ) &=&  \frac{ {d^2\Gamma^{(1/2)}}/{dq^2dm_{K\pi}^2 }- {d^2\Gamma^{(-1/2)}}/{dq^2dm_{K\pi}^2 }}{ {d^2\Gamma}/{dq^2dm_{K\pi}^2 }} \nonumber\\
  &=&  \frac{ \big(-1+ {\hat m_l^2}/{2}\big)( |A_{0}^0|^2 +  |A_{0}^1|^2  + |A_{||}^1|^2 +  |A_{\perp}^1|^2 )   + \frac{3}{2} \hat m_l^2 (|A_{t}^1|^2 + |A_{t}^0|^2 )}{ {d^2\Gamma}/{dq^2dm_{K\pi}^2 }}, 
\end{eqnarray}
and we show the numerical results in   Fig.~\ref{fig:AlambdaBBsKstar}.

\subsection{{\boldmath$B^-\to \pi^+\pi^- \ell\bar\nu$}}

%%%%%%%%%%%%%%%%%%%%%%%%%%%%%%%%%%%%%%%%%%%%%%%%%%%%%%%%%%%%%%%%%%%%%%%%%%%%%%%%
\begin{figure}[ht]
\begin{center}
\includegraphics[scale=0.5]{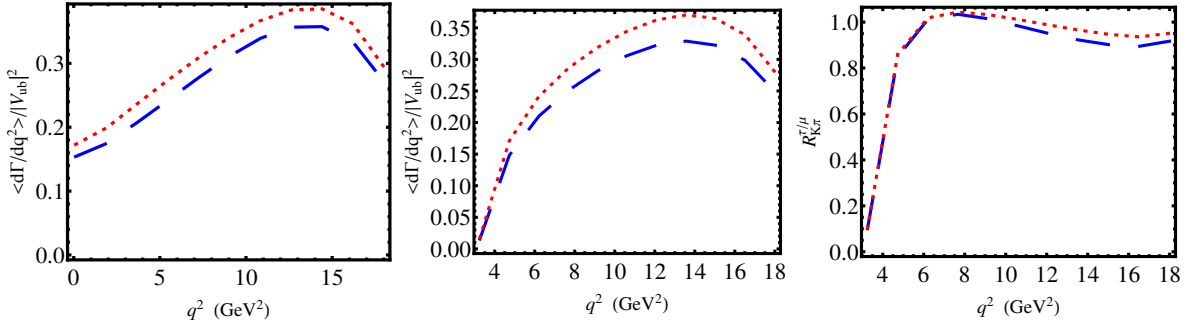}  
\caption{Similar as Fig.~\ref{fig:diffdwBsKstar} but for the   
differential decay widths of the $B^-\to \pi^+\pi^- \ell\bar\nu_{\ell}$.} \label{fig:diffdwBrho}
\end{center}
\end{figure}
%%%%%%%%%%%%%%%%%%%%%%%%%%%%%%%%%%%%%%%%%%%%%%%%%%%%%%%%%%%%%%%%%%%%%%%%%%%%%%%%

%%%%%%%%%%%%%%%%%%%%%%%%%%%%%%%%%%%%%%%%%%%%%%%%%%%%%%%%%%%%%%%%%%%%%%%%%%%%%%%%
\begin{figure}
\begin{center} 
\includegraphics[scale=0.5]{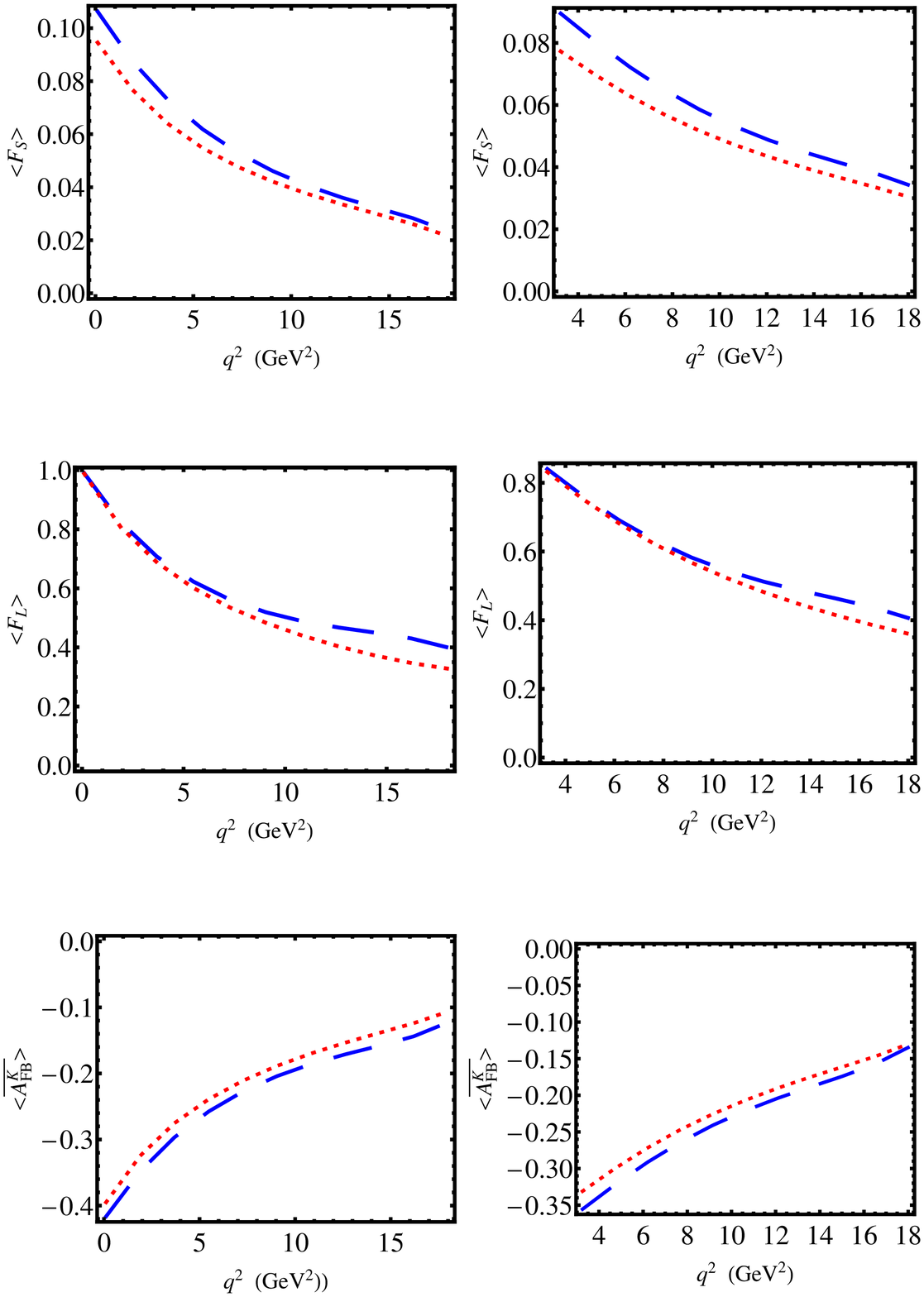}   
\caption{Similar as Fig.~\ref{fig:FSBsKstar} but for the $B^-\to \pi^+\pi^-
  \ell\bar\nu_{\ell}$.} 
\label{fig:FSBrho}
\end{center}
\end{figure}
%%%%%%%%%%%%%%%%%%%%%%%%%%%%%%%%%%%%%%%%%%%%%%%%%%%%%%%%%%%%%%%%%%%%%%%%%%%%%%%% 

In this subsection, we shall  update the predictions for the 
$B^-\to \pi^+\pi^- \ell\bar\nu$ process (for a recent calculation using
dispersion relations matched to chiral perturbation theory, see \cite{HKKMM}).  
For the $B\to \rho$ form factors, 
we take the results from  the LCSR~\cite{Ball:2004rg}  and the LFQM~\cite{Cheng:2003sm} 
calculations. 

We first stress  the importance of  considering the $\pi^+\pi^-$ spectrum  distribution 
and the  $\pi^+\pi^-$ final state interaction. 
To identify the $\rho$ meson, the experimental  measurements from the Babar 
collaboration on $B\to\rho l\bar \nu$~\cite{delAmoSanchez:2010af} have used the 
kinematical constraint  
\begin{eqnarray}
 0.65~ {\rm GeV }< m_{\pi\pi} <0.85~{\rm GeV }~. 
\end{eqnarray}
In the narrow-width limit for the pure P-wave contributions, the integration 
of the lineshape should give
\begin{eqnarray}
 \int dm^2_{\pi\pi}|L_{\rho}(m^2_{\pi\pi})|^2 =1~. 
\end{eqnarray}
However, this integration in the selected  kinematical region amounts to 
\begin{eqnarray}
 \int_{0.65^2}^{0.85^2}dm^2_{\pi\pi}|L_{\rho}(m^2_{\pi\pi})|^2 =0.59,
\end{eqnarray}
which  is far below 1. 
A  smaller value of $|V_{ub}|$ was found by the BaBar 
collaboration~\cite{delAmoSanchez:2010af}:
\begin{eqnarray}
 |V_{ub}|= (2.75\pm 0.24) \times 10^{-3}, 
\end{eqnarray}
based on the data on $B\to\rho l\bar\nu$  in the range $0<q^2< 16~ {\rm GeV}^2$ 
and theoretical results using the LCSR form factors.  Here, theoretical errors
from the $B\to\rho$ form factors are not taken into account in the
experimental analysis~\cite{delAmoSanchez:2010af}. 
This small value is also confirmed by a recent theoretical determination~\cite{Flynn:2008zr}.

 In our calculation, we suggest to choose  $\delta_m =\Gamma_\rho$, corresponding  to 
\begin{eqnarray}
 \int_{(m_\rho-\Gamma_\rho)^2}^{(m_\rho+\Gamma_\rho)^2}dm^2_{\pi\pi}|L_{\rho}(m^2_{\pi\pi})|^2 =0.70. 
\end{eqnarray} 
This  will increase the importance of $B\to \rho \ell\bar\nu$. 
On the other hand  the S-wave lineshape in this region is also significant  
\begin{eqnarray}
 \int_{(m_\rho-\Gamma_\rho)^2}^{(m_\rho+\Gamma_\rho)^2}dm^2_{\pi\pi}|L_{S}(m^2_{\pi\pi})|^2 =0.09.  
\end{eqnarray}
These effects should be taken into account in future experimental determinations.

\begin{figure}
\begin{center} 
\includegraphics[scale=0.6]{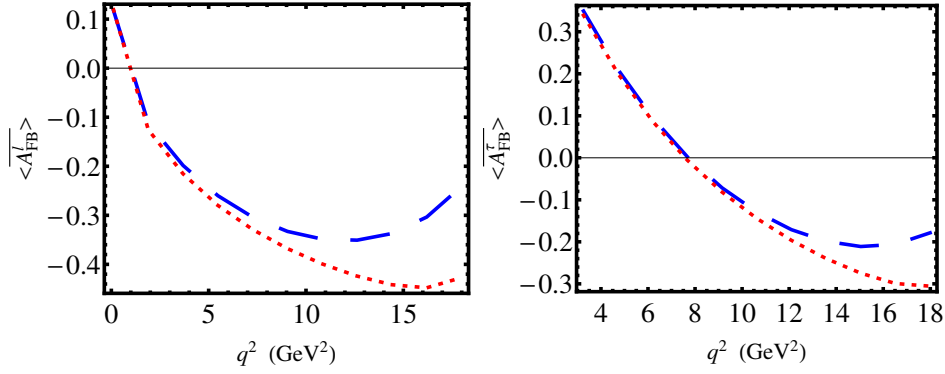}  
\caption{Similar as Fig.~\ref{fig:AFBBsKstar} but for  the
$B^-\to \pi^+\pi^- \ell\bar\nu_{\ell}$ } \label{fig:AFBBrho}
\end{center}
\end{figure}

\begin{figure}
\begin{center}  
\includegraphics[scale=0.5]{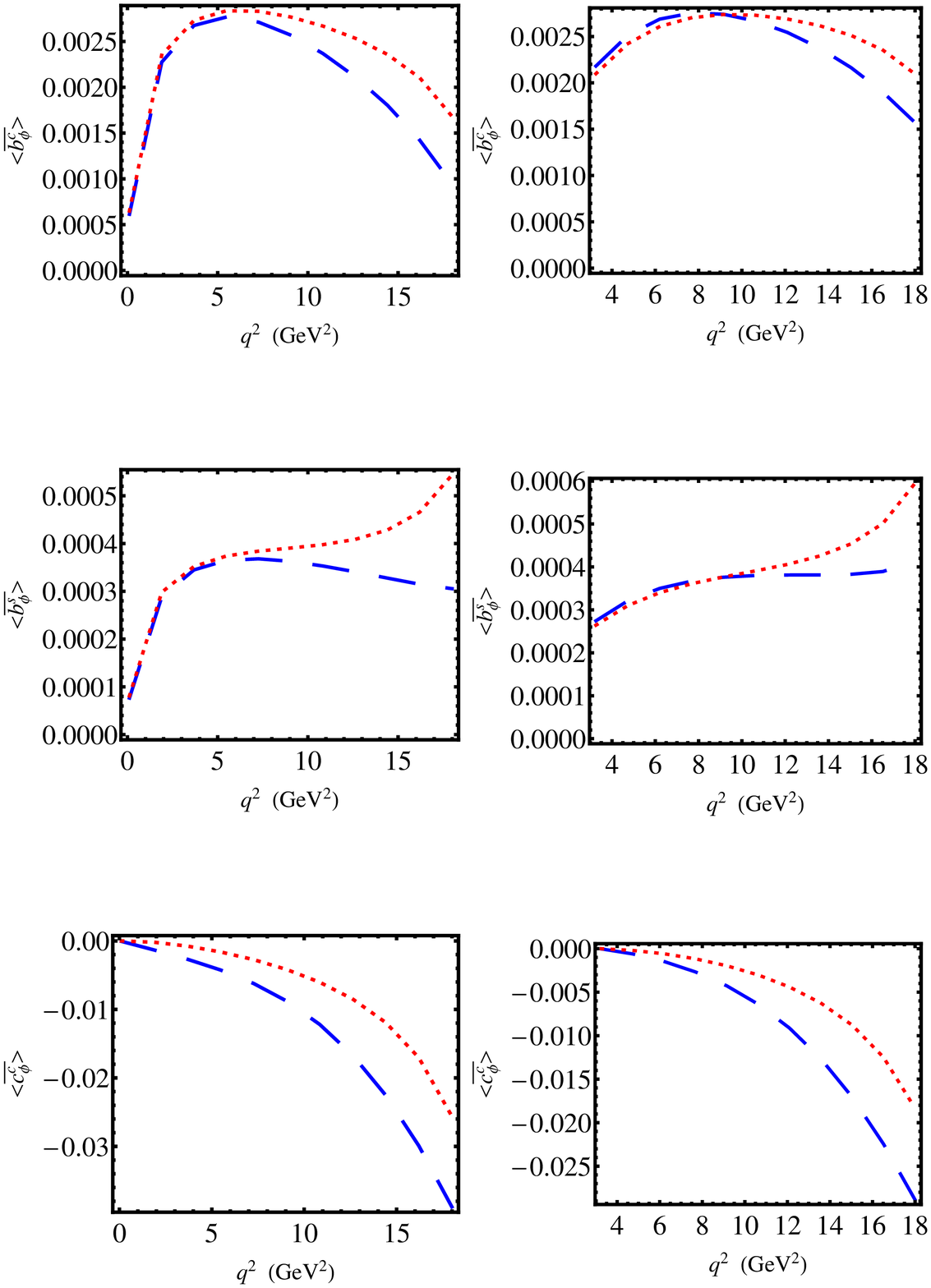} 
\caption{Similar as Fig.~\ref{fig:bcphiBsKstar} but for   the
$B^-\to \pi^+\pi^- \ell\bar\nu_{\ell}$ } \label{fig:bcphiBrho}
\end{center}
\end{figure}

\begin{figure}
\begin{center}
\includegraphics[scale=0.6]{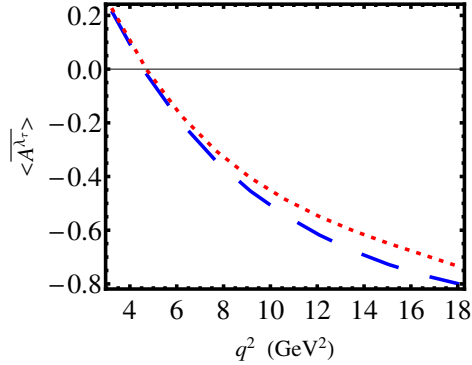}  
\caption{Similar as Fig.~\ref{fig:AlambdaBBsKstar} but for  the
$B^-\to \pi^+\pi^- \ell\bar\nu_{\ell}$} \label{fig:AlambdaBBrho}
\end{center}
\end{figure}

Our results for integrated decay widths  are collected in 
Tab.~\ref{tab:integratedWidthBtopipi}, and for the LFQM calculation, 
we have introduced $10\%$ errors to the form factors, as indicated 
from Refs.~\cite{Chen:2009qk,Wang:2008xt,Wang:2009mi}. The differential decay widths, 
polarisations,  forward-backward asymmetries, and the $\tau$-lepton polarisations 
are given in Figs.~\ref{fig:diffdwBrho}, \ref{fig:FSBrho}, \ref{fig:AFBBrho},  
\ref{fig:bcphiBrho}, and \ref{fig:AlambdaBBrho}, respectively. In these
figures, the dashed and dotted  lines correspond to the LFQM and LCSR form
factors.  The S-wave fraction ranges from approximately  $4\%$ to $10\%$ depending on the $q^2$.

%%%%%%%%%%%%%%%%%%%%%%%%%%%%%%%%%%%%%%%%%%%%%%%%%%%%%%%%%%%%%%%%%%%%%%%%%%%%%%%%
%%%%%%%%%%%%%%%%%%%%%%%%%%%%%%%%%%%%%%%%%%%%%%%%%%%%%%%%%%%%%%%%%%%%%%%%%%%%%%%% 

%%%%%%%%%%%%%%%%%%%%%%%%%%%%%%%%%%%%%%%%%%%%%%%%%%%%%%%%%%%%%%%%%%%%%%%%%%%%%%%%
%%%%%%%%%%%%%%%%%%%%%%%%%%%%%%%%%%%%%%%%%%%%%%%%%%%%%%%%%%%%%%%%%%%%%%%%%%%%%%%% 

%%%%%%%%%%%%%%%%%%%%%%%%%%%%%%%%%%%%%%%%%%%%%%%%%%%%%%%%%%%%%%%%%%%%%%%%%%%%%%%%
%\input{tableBrho.tex}
\begin{table}[t]
\begin{center}
\caption{Integrated decay widths for
 $B^-\to \pi^+ \pi^-  \ell\bar  \nu$:
 $\Delta \zeta_{ \pi \pi}^{\ell} (q_l^2, q_u^2)$. 
Results are given in units of $ps^{-1}$.
The S-wave contributions are obtained 
with form factors calculated in the PQCD approach.}
\label{tab:integratedWidthBtopipi}
\begin{tabular}{cc|cccccccc}
\hline\hline
  & & S-Wave & P-Wave & Total &$f_S(\%)$  \\\hline
$\Delta \zeta_{\pi \pi}^\mu(0, 4)$
 & LFQM  & $0.068\pm 0.038 $ & $ 0.656\pm 0.137 $&$ 0.723\pm 0.176 $&$ 9.41 $\

\\
 & LCSR  & $0.068\pm 0.038 $ & $ 0.754\pm 0.128 $&$ 0.822\pm 0.166 $&$ 8.27 $\

\\
 $\Delta \zeta_{\pi \pi }^\mu(4, 8)$ 
 & LFQM  & $0.068\pm 0.038 $ & $ 0.984\pm 0.206 $&$ 1.05\pm 0.247 $&$ 6.48 $ 
\\
 & LCSR  & $0.068\pm 0.038 $ & $ 1.11\pm 0.205 $&$ 1.18\pm 0.241 $&$ 5.76 $ 
\\
$\Delta \zeta_{\pi \pi }^\mu(8, 12)$
 & LFQM  & $0.064\pm 0.036 $ & $ 1.29\pm 0.268 $&$ 1.35\pm 0.308 $&$ 4.74 $ 
\\
 & LCSR  & $0.064\pm 0.036 $ & $ 1.41\pm 0.261 $&$ 1.47\pm 0.301 $&$ 4.35 $ 
\\
$\Delta \zeta_{\pi \pi }^\mu(12, 16)$
 & LFQM  & $0.054\pm 0.03 $ & $ 1.42\pm 0.294 $&$ 1.47\pm 0.329 $&$ 3.67 $ 
\\
 & LCSR  & $0.054\pm 0.03 $ & $ 1.53\pm 0.297 $&$ 1.58\pm 0.331 $&$ 3.42 $ 
\\
$\Delta \zeta_{\pi \pi }^\mu(0, 16)$
 & LFQM  & $0.254\pm 0.143 $ & $ 4.34\pm 0.916 $&$ 4.6\pm 1.05 $&$ 5.52 $ 
\\
 & LCSR  & $0.254\pm 0.143 $ & $ 4.8\pm 0.895 $&$ 5.06\pm 1.03 $&$ 5.02 $ 
\\
\hline
 $\Delta \zeta_{\pi \pi }^\tau(m_\tau^2, 8)$
 & LFQM  & $0.069\pm 0.039 $ & $ 0.793\pm 0.166 $&$ 0.862\pm 0.205 $&$ 8. $ 
\\
 & LCSR  & $0.069\pm 0.039 $ & $ 0.92\pm 0.171 $&$ 0.988\pm 0.211 $&$ 6.98 $ 
\\
$\Delta \zeta_{\pi \pi }^\tau(8, 12)$
 & LFQM  & $0.074\pm 0.042 $ & $ 1.15\pm 0.241 $&$ 1.22\pm 0.288 $&$ 6.07 $ 
\\
 & LCSR  & $0.074\pm 0.042 $ & $ 1.31\pm 0.246 $&$ 1.38\pm 0.293 $&$ 5.36 $ 
\\
 $\Delta \zeta_{\pi \pi }^\tau(12, 16)$
 & LFQM  & $0.14\pm 0.079 $ & $ 2.43\pm 0.511 $&$ 2.57\pm 0.59 $&$ 5.45 $ 
\\
 & LCSR  & $0.14\pm 0.079 $ & $ 2.76\pm 0.531 $&$ 2.9\pm 0.61 $&$ 4.83 $ 
\\
 $\Delta \zeta_{\pi \pi }^\tau(m_\tau^2, 16)$
 & LFQM  & $0.209\pm 0.117 $ & $ 3.22\pm 0.681 $&$ 3.43\pm 0.797 $&$ 6.09 $ 
\\
 & LCSR  & $0.209\pm 0.117 $ & $ 3.68\pm 0.702 $&$ 3.89\pm 0.818 $&$ 5.37 $ 
\\
\hline
 \hline
\end{tabular} 
\end{center}\end{table}

%%%%%%%%%%%%%%%%%%%%%%%%%%%%%%%%%%%%%%%%%%%%%%%%%%%%%%%%%%%%%%%%%%%%%%%%%%%%%%%%

%%%%%%%%%%%%%%%%%%%%%%%%%%%%%%%%%%%%%%%%%%%%%%%%%%%%%%%%%%%%%%%%%%%%%%%%%%%%%%%%
%%%%%%%%%%%%%%%%%%%%%%%%%%%%%%%%%%%%%%%%%%%%%%%%%%%%%%%%%%%%%%%%%%%%%%%%%%%%%%%% 

\subsection{{\boldmath$B^-\to K^+K^- \ell\bar\nu$} }

For the first time, we provide predictions for  the $B^-\to K^+K^- \ell\bar\nu$ decay rates. 
Apparently, the P-wave resonancde contribution is highly suppressed,  as no 
resonance can be copiously produced in the $b\to u$ transition and has a 
large  decay branching ratio into $K^+K^-$.   The contribution  from 
$\phi$-$\omega$ mixing, whose effects in $B$ decays have been stressed 
in Ref.~\cite{Gronau:2008kk,Li:2009zj,Zhang:2011av}, is given by
\begin{eqnarray}
 {\cal B} (B^-\to (K^+K^-)_P l\bar \nu) \simeq\frac{1}{2} {\cal B} (B^-\to \phi l\bar \nu) = \sin^2\theta  {\cal B} (B^-\to \omega l\bar \nu)  \sim 0.4\times 10^{-6}. 
\end{eqnarray}
where the ${\cal B} (B^-\to \omega l\bar \nu)=(1.19\pm 0.32\pm 0.05)\times 10^{-4} $ 
is taken from Ref.~\cite{Adachi:2008kn} and the 
$\omega-\phi$ mixing   angle $\theta=(3.4\pm 0.3)^\circ$ is  extracted from 
the $\omega$ and $\phi$ decays in Ref.~\cite{Kucukarslan:2006wk}.

It is also necessary to consider  the $D$ wave contributions from the $f_2(1270)$
\begin{eqnarray}
 {\cal B} (B^-\to (K^+K^-)_Dl\bar \nu) = 4.6\%\times 0.52\times 10^{-4} = 2.3\times 10^{-6},
\end{eqnarray}
where we have used the estimate of the ${\cal B}(B\to f_2 \ell\bar\nu)$ from Ref.~\cite{Wang:2010ni}.

The S-wave $K^+K^-$ can have isospin 0 and isospin 1, and here we focus on the isospin 0 contribution.   
The $f_0(980)$ contribution is suppressed due to phase space. Integrating 
from the threshold to the $f_2(1270)$ mass, we have 
\begin{eqnarray}
 \int_{4m_{K}^2}^{m_{f_2(1270)}^2}  dm_{K\bar K}^2  |L_{S}(m_{K\bar K}^2)| = 0.8\%,
\end{eqnarray}
and thus  
\begin{eqnarray}
 {\cal B} (B^-\to (K^+K^-)_Sl\bar \nu) = 0.8 \%\times 1.8\times 10^{-4} = 1.4\times 10^{-6},
\end{eqnarray}
where the transition induced by  $b\to u\ell\bar\nu$ is calculated using 
the $B\to a_0(980)$ form factors. 
The measurements of this channel at the  LHCb  facility and the Super B factory in the
future  may directly test our mechanism and constrain the parameters in this approach.

%%%%%%%%%%%%%%%%%%%%%%%
\section{Conclusions}
\label{sec:conclusions}
%%%%%%%%%%%%%%%%%%%%%%%

In this work,  we have  analysed  the $\overline B_s^0\to K^+l^-\bar \nu$ 
and $\overline B_s^0\to K^{*+}\ell^-\bar \nu$ decays in the pursuit of the 
extraction  of the CKM matrix element $|V_{ub}|$.  We have calculated 
differential and  integrated decay widths  in units of $|V_{ub}|^2$ based on  
three sets of  $B_s\to K$ form factors. Although parametric errors in the decay 
widths will drop out in ratios, the three sets of results have different 
behaviours especially  in the large $q^2$ region. Such discrepancies will be 
removed once  Lattice QCD  is able to predict the form factors at  large $q^2$.

For the decay $B_s\to K\pi l\bar\nu$,  we have derived the angular distributions 
with the inclusion of the S-wave $K\pi$ contributions, from which a number of  
angular observables including partial widths, polarisation, or the forward-backward 
asymmetry,  can be defined.  Using the recent Lattice QCD   and the light-cone 
sum rule results for the $B_s\to K^*$ form factors,  we have discussed the 
possible S-wave effects on angular distributions. We  found that the S-wave 
contribution to partial widths  can reach $10\%$ to $20\%$ depending on the 
momentum transfer, and thus their impact can not be ignored.  We have  also 
updated the results on  $B\to \pi^+\pi^-\ell\bar\nu$ and  discussed the S-wave 
$\pi\pi$ contributions.  Measurements of these channels in the future can not 
only be used to extract $|V_{ub}|$ but also  provide  useful information to access 
the $K\pi/\pi\pi$  strong phase.

\section*{Acknowledgements}
We thank Michael D\"oring for collaboration in the early stage of this work,
and Vladimir Galkin and Bastian Kubis for  useful discussions.   
This work is supported in
part by the DFG and the NSFC through funds provided to the Sino-German CRC 110
``Symmetries and the Emergence of Structure in QCD'', and the EU ``I3HP Study of
Strongly Interacting Matter''  under the Seventh Framework Program.

\begin{appendix}
\section{Angular coefficients} 

Substituting the  expressions $A_i$ into the angular coefficients, 
we obtain the general  expressions
\begin{eqnarray}
 I_1	 &=& 
 \sum_{J=0,...}  \bigg\{ |Y_J^0(\theta_K, 0)|^2 \left[(1+\hat m_l^2) |A^J_{0}|^2  
 +2 \hat m_l^2  |A_t^J|^2\right]   \nonumber\\
 &&+ 2\sum_{ J'=J+1, ...} Y_J^0(\theta_K, 0)Y_{J'}^0(\theta_K, 0) \left[ \cos(\delta_{0}^J - 
 \delta^{J'}_{0})|A^J_{0}||A^{J'*}_{0}|   + 2\hat m_l^2 
 \cos (\delta_{t}^J -\delta_{t}^{J'} )|A^J_t||A^{J'}_t|\right]\bigg\}
 \nonumber\\
 & &  +  \frac{3+\hat m_l^2}{2}  \sum_{J=1,...}  \bigg\{ |Y_J^{-1}(\theta_K, 0)|^2 \left[   [|A^J_{\perp}|^2+|A^J_{||}|^2 ]  \right] \nonumber\\
 && +  \sum_{ J'=J+1, ...} Y_J^{-1}(\theta_K, 0)Y_{J'}^{-1}(\theta_K, 0) \left[    2\cos(\delta_{\perp}^{J} 
 - \delta_{\perp}^{J'})|A_{\perp}^J||A_{\perp}^{J'}| \right]  \bigg\},\nonumber\\
% \end{eqnarray} 
 %%%%
% \begin{eqnarray}
 I_2  &=& -\beta_l   \sum_{J=0,...}  \bigg\{ |Y_J^0|^2   |A^J_{0}(\theta_K, 0)|^2   + 2\sum_{ J'=J+1, ...} Y_J^0(\theta_K, 0)Y_{J'}^0(\theta_K, 0)
  \cos(\delta_{0}^J - \delta^{J'}_{0})|A^J_{0} A^{J'}_{0}|  \bigg\} \nonumber\\
% \end{eqnarray} 
 %%%%
% \begin{eqnarray}
 %%%%%%%%%
  & &+ 
 \frac{1}{2}\beta_l  \sum_{J=1,...}  \bigg\{  |Y_J^{-1}(\theta_K, 0)|^2  (|A^J_{\perp}|^2+|A^J_{||}|^2) \nonumber\\
 &&       +2\sum_{ J'=J+1} Y_J^{-1}(\theta_K, 0)Y_{J'}^{-1}(\theta_K, 0)\left[ \cos(\delta_{\perp}^J 
 - \delta^{J'}_{\perp})|A^J_{\perp} A^{J'}_{\perp}| + \cos(\delta_{||}^J 
 - \delta^{J'}_{||})|A^J_{||}A^{J'}_{||}| \right] \bigg\},
 \nonumber\\
% \end{eqnarray} 
 %%%%
% \begin{eqnarray}
 %%%%%%%%%
 I_3   &=& \beta_l \sum_{J=1,...}  \bigg\{  |Y_J^{-1}(\theta_K, 0)|^2  (|A^J_{\perp}|^2-|A^J_{||}|^2)         \nonumber\\
 && +2\sum_{ J'=J+1} Y_J^{-1}(\theta_K, 0)Y_{J'}^{-1}(\theta_K, 0)\left[ \cos(\delta_{\perp}^J 
 - \delta^{J'}_{\perp})|A^J_{\perp} A^{J'}_{\perp}| - \cos(\delta_{||}^J 
 - \delta^{J'}_{||})|A^J_{||}A^{J'}_{||}| \right] \bigg\},
 \nonumber
 \end{eqnarray} 
 %%%%
 \begin{eqnarray}
 %%%%%%%
 I_4 
  &=&2 \beta_l   \sum_{J=0, ...} \sum_{J'=1, ..} \left[  Y_J^0(\theta_K, 0) Y_{J'}^{-1}(\theta_K, 0) | A^J_{0}A^{J'*}_{||} | \cos(\delta_{0}^J -\delta_{||}^{J'})  \right],
  \nonumber\\
% \end{eqnarray} 
 %%%%
% \begin{eqnarray}  %%%%%%%%%%%
 I_5 
  &=&4   \sum_{J=0, ...} \sum_{J'=1, ..}Y_J^0(\theta_K, 0) Y_{J'}^{-1} (\theta_K, 0) \left[   | A^J_{0}A^{J'*}_{\perp} | \cos(\delta_{0}^J -\delta_{\perp}^{J'}) -\hat m_l^2  | A^J_{t}A^{J'*}_{||} | \cos(\delta_{t}^J -\delta_{||}^{J'})\right],
  \nonumber\\
% \end{eqnarray} 
 %%%%
% \begin{eqnarray}
  %%%%%%%%%
 I_6  &=& 4 \sum_{J,J'=1,...}  \bigg\{   Y_J^{-1} (\theta_K, 0)Y_{J'}^{-1} (\theta_K, 0)| A^J_{||}A^{J'*}_{\perp} | \cos(\delta_{||}^J -\delta_{\perp}^{J'})  \bigg\}   \nonumber\\
 && +  \hat m_l^2 \sum_{J,J'=0,...}  \bigg\{   Y_J^{0}(\theta_K, 0) Y_{J'}^{0}(\theta_K, 0) | A^J_{t}A^{J'*}_{0} | \cos(\delta_{t}^J -\delta_{0}^{J'})  \bigg\},
 \nonumber\\
% \end{eqnarray} 
 %%%%
% \begin{eqnarray}
  I_7 
 &=&4 \sum_{J=0, ...} \sum_{J'=1, ..} Y_J^0(\theta_K, 0) Y_{J'}^{-1}(\theta_K, 0)  \left[  | A^J_{0}A^{J'*}_{||} | \sin(\delta_{0}^J -\delta_{||}^{J'}) -\hat m_l^2 | A^J_{t}A^{J'*}_{\perp} | \sin(\delta_{t}^J -\delta_{\perp}^{J'})\right],
 \nonumber\\
% \end{eqnarray} 
 %%%%
% \begin{eqnarray}
 %%%%%%%
 I_8 &=&
 2 \beta_l  \sum_{J=0, ...} \sum_{J'=1, ..} \left[  Y_J^0 (\theta_K, 0)Y_{J'}^{-1}(\theta_K, 0) | A^J_{0}A^{J'*}_{\perp} | \sin(\delta_{0}^J -\delta_{\perp}^{J'})  \right],
 \nonumber\\
% \end{eqnarray} 
 %%%%
% \begin{eqnarray}
 %%%%%%%%
 I_9
 &=& 2\beta_l \sum_{J=1, ...} \sum_{J'=1, ..} \left[  Y_J^{-1}(\theta_K, 0) Y_{J'}^{-1}(\theta_K, 0) | A^J_{\perp}A^{J'*}_{||} | \sin(\delta_{\perp}^J -\delta_{||}^{J'})  \right].  
\end{eqnarray}

\end{appendix}

%%%%%%%%%%%%%%%%%%%%%%%%%%%%%%%%%%%%%%%%%%%%%%%%%%%%%%%%%%

\end{document}